# Graphene oxide upregulates the homeostatic functions of primary astrocytes and modulates astrocyte-to-neuron communication

*Martina Chiacchiaretta[†°], Mattia Bramini[†°], Anna Rocchi[†], Andrea Armirotti[#], Emanuele Giordano[†], Ester Vázquez[‡], Tiziano Bandiera[#], Stefano Ferroni[§], Fabrizia Cesca[†*ϕ] and Fabio Benfenati[†*ϕ]*

[†]Center for Synaptic Neuroscience and Technology and Graphene Labs, Istituto Italiano di Tecnologia, 16163 Genova, Italy; [#]Drug Discovery and Development and Graphene Labs, Istituto Italiano di Tecnologia, 16163 Genova, Italy; [‡]Departamento de Química Orgánica, Universidad de Castilla La-Mancha, 13071 Ciudad Real, Spain; [§]Department of Pharmacy and Biotechnology, University of Bologna, 40126 Bologna; *IRCCS Ospedale Policlinico San Martino, Genova, Italy.

°Equal contribution
[ϕ] Senior authors

*Corresponding authors:*
Mattia Bramini, PhD; e-mail: mattia.bramini@iit.it
Stefano Ferroni, PhD; e-mail: stefano.ferroni@unibo.it

**ABSTRACT**

Graphene-based materials are the focus of intense research efforts to devise novel theranostic strategies for targeting the central nervous system. In this work, we have investigated the consequences of long-term exposure of primary rat astrocytes to pristine graphene (GR) and graphene oxide (GO) flakes. We demonstrate that GR/GO interfere with a variety of intracellular processes as a result of their internalization through the endo-lysosomal pathway. Graphene-exposed astrocytes acquire a more differentiated morphological phenotype associated with extensive cytoskeletal rearrangements. Profound functional alterations are induced by GO internalization, including the upregulation of inward-rectifying $K^+$ channels and of $Na^+$-dependent glutamate uptake, which are linked to the astrocyte capacity to control the extracellular homeostasis. Interestingly, GO-pretreated astrocytes promote the functional maturation of co-cultured primary neurons by inducing an increase in intrinsic excitability and in the density of GABAergic synapses. The results indicate that graphene nanomaterials profoundly affect astrocyte physiology *in vitro,* with consequences for neuronal network activity. This work supports the view that GO-based materials could be of great interest to address pathologies of the central nervous system associated to astrocyte dysfunctions.

There is a growing interest on the potential use of graphene (G)-based nanomaterials for biomedical applications directed to the central nervous system (CNS). To meet this challenge, thorough investigations on the safety and biocompatibility of G nanostructures in contact with the diverse brain cell types are required. This issue is extremely relevant for G materials, since their concentration, lateral dimension, charge and surface structure are known to influence their bio-interactions and need to be investigated in detail.[1, 2] G-based scaffolds were shown to support the survival, growth and differentiation of primary neurons, thus representing safe and promising tools to promote neural growth and regeneration *in vivo*.[3-6] However, recent studies reported that acute and chronic exposure of primary hippocampal and cortical neurons to graphene oxide (GO) caused altered $Ca^{2+}$ dynamics and an excitatory/inhibitory imbalance in favor of inhibitory synaptic transmission.[7, 8]

Astrocytes, which are the most numerous cell population in the mammalian CNS, are involved in the structural and functional regulation of neuronal circuits and contribute to maintain the homeostasis of the perineuronal *milieu* through the expression of specific ion channels and transporters.[9] Astrocytes respond to pathological insults *in vivo* by changing some of their molecular and functional features, a process called 'reactive astrocytosis',[10] which can be beneficial or detrimental to the integrity and functionality of neuronal circuits depending on the specific pathological setting.[11] The paramount importance of astrocytes in CNS physiology prompted us to address the impact of GR/GO exposure on this cell type, with the aim of providing further information on the biocompatibility profile of these nanomaterials when applied to the CNS and on their potential applications to treat astrocyte-related pathologies.

Very limited data on the impact of G-based nanomaterials on astrocytes are presently available. G-films were able to accelerate the maturation of neural stem cells and their progenies, including glial cells, by affecting their active and passive bioelectric membrane properties.[12] Moreover, GO nanosheets did not alter astrocyte viability *in vitro*, but affected their ability to release synaptic-like microvesicles involved in neuron-astrocyte communication.[7] PEGylated rGO caused morphological changes and ROS production, and affected astrocyte viability *in vitro* and *in vivo*.[13] Although these studies highlighted specific aspects of the interaction between G materials and astrocytes, no information is available to date on the molecular, cellular and functional consequences of exposing astrocytes to G nanomaterials.

In this work, we carried out a detailed investigation of the effects of GR/GO exposure on the functional properties of primary rat cortical astrocytes. Although GR and GO did not affect cell viability, they promoted a marked change in cell shape from an epithelioid morphology to an asymmetrical shape with elongated processes, which was associated with extensive cytoskeletal rearrangements. Exposure to GO, but not to GR, caused the upregulation of $K^+$ channels involved in extracellular $K^+$ homeostasis and an increase in glutamate clearance that were causally related to GO internalization. Interestingly, GO-treated astrocytes affected the functional properties of co-cultured neurons, by potentiating inhibitory synaptic transmission and accelerating the maturation of intrinsic neuronal excitability. These results indicate that GO profoundly affects astrocyte physiology *in vitro*, with repercussions on neuronal network activity. This work highlights the intrinsic different biological effects elicited by different nanomaterials and supports the relevance of GO for future applications in CNS pathologies associated to astrocyte dysfunctions.

**RESULTS AND DISCUSSION**

*GR and GO induce morphological changes without affecting astrocyte viability*

We characterized the physiological response of primary astrocytes to prolonged exposure to G-based materials, namely few-layer GR and GO. The former was prepared by exfoliation of graphite through interaction with melamine by ball-milling treatment, as described by Leon *et al.*[14] GO, provided by Grupo Antolin Ingeniería (Burgos, Spain), was obtained by oxidation of carbon fibers. A complete description of material preparation and physical-chemical characterization is reported in our previous work.[8]

We isolated primary astrocytes from neonatal rat brain cortices (**Figure S1**) and cultured them for up to 7 days *in vitro*. Twenty-four hours after plating, cultures were exposed to an aqueous dispersion of 10 µg/ml of either GR or GO for 24 h, 72 h and 7 days. The viability of G- and vehicle-treated cultures (Ctrl: 0.09 ppm melamine/$H_2O$ for GR or $H_2O$ for GO) was evaluated as a function of the exposure time using immunohistochemistry and flow cytometry of propidium iodide (PI) to detect apoptotic cells. Exposure to G materials was not harmful to primary astrocytes at any of the tested exposure times (**Figure S2a,b**), indicating that GR and GO do not affect astrocyte survival nor induce astrogliopathy.[15]

We next evaluated whether exposure to GR/GO altered cell morphology. To this end, astrocytes were immunostained with antibodies to the intermediate filament glial fibrillary acidic protein (GFAP) or subjected to scanning electron microscopy (SEM). Immunofluorescence revealed that exposure to G-materials induced a clear shape change that, from the regular and flat shape typical of quiescent primary astrocytes, became irregular and characterized by multiple thin processes and elongated protrusions resembling the morphology of *in vitro* activated/mature glia.[16, 17] We quantitatively analyzed this phenomenon from SEM images using the circularity index as a measure of the symmetry and regularity of the cell shape. Upon exposure to either GR or GO, astrocytes are characterized by a markedly lower index, indicative of a more elongated and asymmetrical shape with respect to vehicle-treated cells (**Figure 1a**).

To better investigate the physical interaction of GR and GO with astrocytes, transmission electron microscopy (TEM) was used. As shown in **Figure 1b**, both GR and GO flakes (black arrowheads) were present inside the cells as aggregates of various sizes and compactness, mainly localized within membrane-bound vesicles, likely belonging to the endo-lysosomal pathway. The non-homogenous flake dispersion in solution may explain the formation of aggregates, although we cannot exclude that these agglomerates are formed inside cells upon fusion of multiple vesicles into larger intracellular structures.

To get more insights into the process of G internalization and identify the intracellular organelles containing the internalized flakes, we focused on the endo-lysosomal pathway, known to be the preferential destination of nanoparticles and nanomaterials.[18-21] Astrocytes exposed to G-flakes for 24 h, 72 h and 7 days were labeled with antibodies to early-endosomal (EEA1) and lysosomal (LAMP1) markers. Three-dimensional Z-stack images were acquired with the reflection light acquisition modality,[8, 22] which allows the visualization of G-flakes also in confocal mode and the precise quantification of their uptake and intracellular location (**Figure S2c,d**). When the percentage of GR/GO flakes overlapping with the cell area was quantified, the vast majority of astrocytes were found to internalize GR/GO, with percentages of internalized flakes over total flakes already very high after 24 h of exposure (**Figure 1c**; GR 42 ± 2.2%; GO 35 ± 1.1%). The co-localization of intracellular G particles with EEA1 and LAMP1 revealed that GR/GO flakes were initially internalized by EEA1-positive early endosomes and underwent a time-dependent transition to LAMP1-positive organelles in which they reached the highest value at 7 days (**Figure 1d**), in line with previous studies describing the intracellular trafficking of nanomaterials in living systems.[19-

[21, 23] In addition, the expression and maturation of LC3, a marker of autophagosomal membranes, investigated by immunofluorescence and western blotting (**Figure S3**), showed a complete absence of LC3 activation (LC3 II/I transition) in both GR- and GO-treated astrocytes, in contrast with the strong autophagy reaction observed in primary neurons upon G-flakes exposure.[8]

The observed morphological changes could be due to G-induced cytoskeletal rearrangements, and/or to signaling cascades initiated by the physical interaction of G flakes with the cytoskeleton, phenomena that have been reported in other cellular models.[24, 25] Differently from carbon nanotubes, which have direct effects on GFAP,[26] G flakes, due to their size and shape, might be able to insert into the inter-strand gap of the actin tetramer and cause its dissociation into monomers, which could eventually lead to the disruption of actin filaments.[24] To explore this issue, astrocytes exposed to either GR or GO for 72 h were stained with fluorescent phalloidin and anti-α-tubulin antibodies to visualize actin filaments and microtubules, respectively. Three-dimensional Z-stack confocal images were acquired by adopting the reflection light acquisition modality to concomitantly visualize G-flakes.[8, 22] While control cells showed well-organized actin and tubulin filament bundles, both GR- and GO-treated astrocytes displayed breakdown of actin fibers and rearrangement of α-tubulin that irregularly aggregated upon G exposure (**Figure 1e**). These data suggest that the observed morphological changes are linked to the disruption/reorganization of the astrocyte cytoskeleton in response to G-flakes internalization.

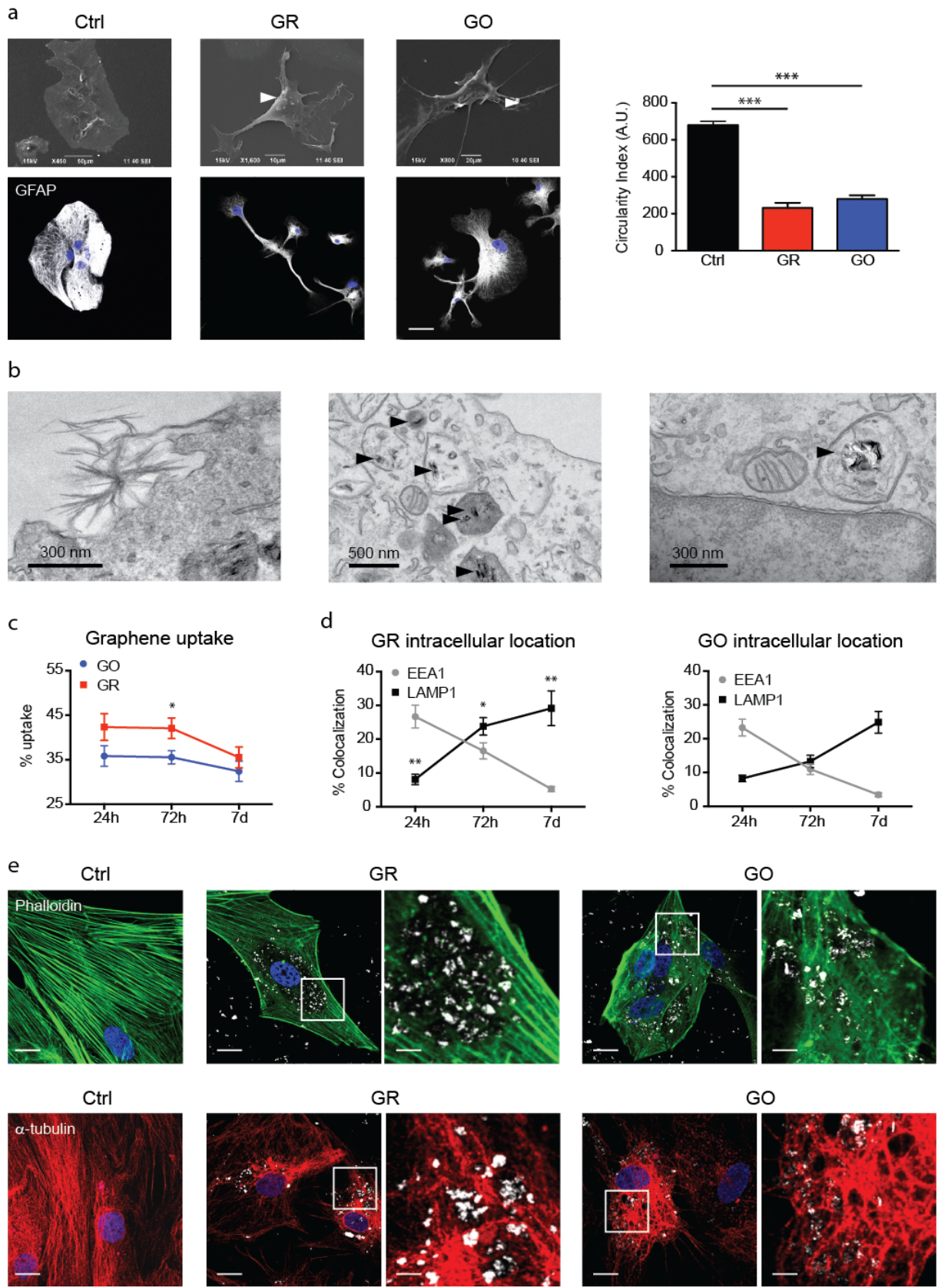

**Figure 1. Interactions of GR and GO with primary astrocytes.** (**a**) Astrocytes exposed to GR and GO for 72 h were examined by SEM and immunofluorescence with anti-GFAP antibodies (white) and Hoechst33342 (blue). *Left:* The morphology of cells exposed to GR and GO was severely

affected compared to untreated samples (Ctrl). Scale bar, 50 μm. *Right:* Circularity index calculated from SEM images for each experimental condition (n = 100 cells, from 3 independent cell preparations). (**b**) The cell uptake of GR and GO and the intracellular location of G-flakes were confirmed by TEM. Large flakes were found outside the cells in close proximity to the cell membrane (*left*). Large amounts of G-flakes (black arrowheads) were observed into intracellular vesicles, such as lysosomes (*middle*) and larger lipid vesicles (*right*). (**c**) Total internalization of GR and GO was quantified over time and expressed as percent uptake with respect to the total G. (**d**) The extent of co-localization of GR (*left*) and GO (*right*) with EEA1 and LAMP1 positive organelles was analyzed using the JACOP plugin of the ImageJ software (50 cells from n = 3 coverslips per experimental condition, from 3 independent cell preparations). (**e**) Confocal images of astrocytes stained with phalloidin (green, *upper panel*) and anti-α-tubulin (red, *lower panel*) to visualize actin filaments and microtubules, respectively, and with Hoechst33342 (blue) for cell nuclei. Scale bars: 5 and 1 μm for low and high magnification, respectively. Data are shown as means ± sem. *$p<0.05$, **$p<0.01$, ***$p<0.001$, one-way ANOVA/Bonferroni's tests.

## *Differential effects of GR and GO on astrocyte electrical membrane properties*

Given the observed morphological changes, we assessed whether G-flakes affected the electrical membrane properties of astrocytes by performing patch-clamp recordings of $K^+$ currents before and after exposure to either GR or GO (1-10 μg/ml; 72 h). It is well known that primary astrocytes mainly express outward rectifying $K^+$ channels, and that only upon functional differentiation they acquire a more physiological $K^+$ current profile characterized by the expression of inward rectifier $K^+$ (Kir) channels.[27, 28] Astrocytes were voltage-clamped at a holding potential of -60 mV and challenged with voltage ramps or step protocols. Typical current traces elicited by both protocols in astrocytes exposed to GR are shown in **Figure 2a**. Treatment with GR did not significantly affect the amplitude and slope of the ramp currents; the current density measured at 120 mV was not significantly different between GR-treated and vehicle-treated samples (25.1 ± 4.8 pA/pF for Ctrl, 29.4 ± 6.7 pA/pF for GR). Addition of barium ($Ba^{2+}$, 100 μM), which predominantly blocks inwardly rectifying $K^+$ channels,[29] slightly inhibited (~ 20%) the current response at 120 mV to a similar extent in vehicle- and GR-treated astrocytes. To unveil the time-dependent kinetics of the elicited current, we applied a step protocol. Upon step depolarization, rapidly activating and non-inactivating outward currents typical of delayed rectifier $K^+$ conductance were elicited at potentials > -20 mV. The voltage and kinetics of the currents were not significantly different in vehicle- and GR-treated cells and the analysis of the current-voltage relationships depicted a similar voltage-dependent profile between vehicle- and GR-treated astrocytes (**Figure 2b**). Compared to vehicle-treated astrocytes, GR exposure did not affect the resting membrane potential (-28.0 ± 6.0 mV for Ctrl, -35.1 ± 5.2 mV for GR), the input resistance (686 ± 51 MΩ for Ctrl, 756 ± 133 MΩ for GR) or the specific conductance (0.037 ± 0.005 nS/pF for Ctrl, 0.038 ± 0.005 nS/pF for GR) (**Figure 2c**).

By contrast, GO-treated astrocytes displayed a marked increase in outward rectifying currents compared to vehicle-treated cells (**Figure 2d**). Analysis of current density values recorded at 120 mV indicated an almost three-fold increase in GO-treated cells (26.9 ± 5.4 pA/pF for Ctrl, 67.1 ± 11.4 pA/pF for GO). Moreover, administration of $Ba^{2+}$ produced a significantly larger (~ 40%) inhibition of the ramp current measured at 120 mV compared to vehicle-treated astrocytes. The time and voltage-dependent kinetics of the current evoked with the voltage stimulation protocol were similar under the two experimental conditions. The current-voltage relationship shows that the increase in current amplitude observed in GO-treated astrocytes started at negative potentials and became more pronounced at positive membrane potentials (**Figure 2e**). Furthermore, GO hyperpolarized the resting membrane potential (-30.9 ± 6.5 mV for Ctrl and -50 ± 8 mV for GO), induced a significant decrease in input resistance (706 ± 50 MΩ for Ctrl, 487 ± 67 MΩ for GO) and augmented the specific conductance (0.03 ± 0.003 nS/pF for Ctrl, 0.06 ± 0.01 nS/pF for GO) (**Figure 2f**). Interestingly, a ten-fold lower dose of GO (1 μg/ml) elicited similar changes in specific conductance and current density, with a trend toward hyperpolarization and input resistance increase, revealing that the GO modulation is specific and dose-dependent (**Figure S4**). These data suggest that $Ba^{2+}$-sensitive, weakly inward-rectifying Kir channels, which are typically upregulated in differentiated cultured astrocytes [28, 30] and downregulated in several glial-related pathologies,[31] are likely responsible for mediating the increase in current amplitudes and the variations in the passive membrane properties of astrocytes in response to GO exposure. These findings extend previous reports in which other carbon-based materials, namely single-wall (SWCNTs) and multi-wall (MWCNTs) carbon nanotubes, have been shown to interact with $K^+$ channels and alter their physiological functions.[32-34]

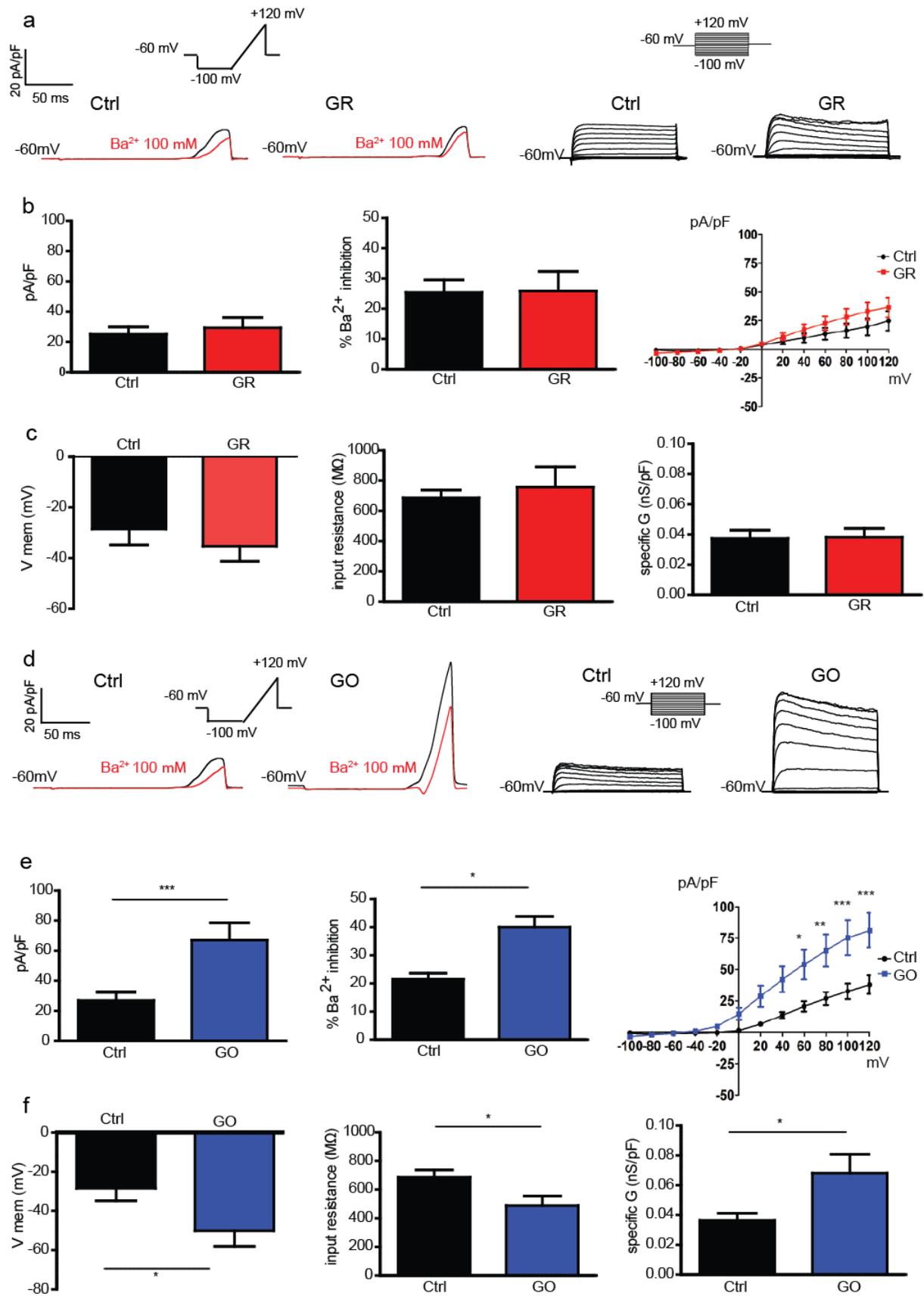

**Figure. 2. Exposure to GO, but not GR, altered specific electrophysiological properties of astrocytes.** (**a**) Whole-cell patch-clamp recordings of primary astrocytes exposed to GR. *Left:* Representative current traces elicited with a voltage ramp protocol in astrocytes treated for 72 h with either GR (10 μg/ml) or vehicle before (black traces) or after the addition of $Ba^{2+}$ (100 μM, red traces) to the recording medium. *Right:* Representative traces of whole-cell currents evoked with a family of voltage steps. (**b**) GR exposure does not affect outward currents in astrocytes. *Left:* Current density values at 120 mV (n = 19 cells per experimental condition). *Middle:* Percentage of current inhibition at 120 mV in the presence of $Ba^{2+}$ (n = 10 cells per experimental condition). *Right:* Current-voltage relationships in astrocytes exposed to GR (red) or vehicle (black) (n = 12 cells per experimental condition). (**c**) Passive membrane properties of astrocytes treated for 72 h with either GR (10 μg/ml) or vehicle. The resting membrane potential (mV; *left*), input resistance (MΩ; *middle*) and specific conductance (nS/pF; *right*) were unchanged (n = 19 cells per experimental condition). (**d**) Whole-cell patch-clamp recordings of primary astrocytes exposed to GO. *Left:* Representative current traces elicited with a voltage ramp protocol in astrocytes treated for 72 h with either GO (10 μg/ml) or vehicle before (black traces) or after exposure to $Ba^{2+}$ (100 μM, red traces) to the recording medium. *Right:* Representative traces of whole-cell currents evoked with a family of voltage steps. (**e**) GO exposure markedly increases outward $K^+$ conductance in astrocytes. *Left:* Current density values at 120 mV (n = 19 cells per experimental condition). *Middle:* Percentage of current inhibition at 120 mV after application of 100 μM $Ba^{2+}$ (n = 10 cells per experimental condition). *Right:* Current-voltage relationships in astrocytes exposed to GO (blue) or vehicle (black) (n = 12 cells from per experimental conditions). (**f**) GO exposure for 72 h affects the passive membrane properties of astrocytes. Resting membrane potential (mV; *left*), mean input resistance (MΩ; *middle*) and specific conductance (nS/pF; *right*) are shown (n = 19 cells per experimental condition). All data are collected from 3 independent cell preparations and are expressed as mean ± sem. *p<0.05, ***p<0.001, two-tailed Student's *t*-test/ Mann-Whitney *U*-test.

***The effects of GO on astrocyte membrane properties depend on the increased expression of Kir4.1 channels***

We next investigated whether the expression of Kir4.1 channels, the main component of the astroglial Kir current involved in $K^+$ buffering,[10] was affected by G exposure. Primary astrocytes treated with either GR or GO were stained with anti-Kir4.1 antibodies by avoiding the permeabilization step in the immunostaining procedure, which allowed monitoring the expression of Kir4.1 channels at the plasma membrane. Membrane labeling of Kir4.1 was higher and more diffuse in GO-treated astrocytes, whereas in GR-treated cells it was indistinguishable from control astrocytes (**Figure 3a**). To confirm the immunofluorescence results, G-treated and control astrocytes were subjected to western blotting. The total cell expression of Kir4.1 was also selectively increased upon exposure to GO, with a significant upregulation of both the monomeric and tetrameric forms of the channel (**Figure 3b**), which were instead not altered by GR (**Figure S5**). These data confirm that the changes in current amplitudes and passive membrane properties observed in GO-treated astrocytes can be attributed to the enhanced expression of Kir4.1 channels.

Kir4.1 channels control extracellular homeostasis also through the regulation of glutamate dynamics. In rodents, the expression of Kir4.1 has been linked to a positive modulation of glutamate clearance by astrocytes mediated by $Na^+$-dependent uptake through glutamate transporters.[35, 36] To address the possibility that the increase in Kir4.1 current could in turn increase glutamate clearance, we evaluated $^3$H-labeled glutamate ($^3$H-Glu) uptake in primary astrocytes (**Figure 3c**). The cells were exposed to GR or GO for 72 h and then challenged with $^3$H-Glu in the presence or absence of choline in the extracellular bath, to isolate the $Na^+$-dependent glutamate transport or following administration of $Ba^{2+}$ to inhibit Kir4.1 channels. As expected, while GR did not promote any change in glutamate clearance compared to control, GO-treated astrocytes showed a marked increase in the uptake of extracellular glutamate. Importantly, replacement of $Na^+$ with choline or $Ba^{2+}$ decreased glutamate uptake similarly to what observed upon treatment with the competitive blocker of glutamate transporters dl-threo-β-benzyloxyaspartic acid (DL-TBOA; 100 μM),[37] indicating the strong contribution of Kir4.1 activity to the $Na^+$-dependent uptake of glutamate (**Figure 3c**).

We next explored whether the increase in glutamate uptake was associated with the enhanced expression of the two specific astroglial transporters GLT-1 and GLAST.[38] Immunofluorescence analysis of membrane-targeted transporters revealed a selective upregulation of GLT-1 membrane expression in GO-treated astrocytes (**Figure 3d**) compared to GR- or vehicle-treated astrocytes, while the membrane-targeted pool of GLAST transporters was not affected (**Figure 3f**). Western blotting experiments confirmed that total GLT-1 and GLAST expression was comparable under all the experimental conditions (**Figure 3e,g**). These data support the view that the long-term challenge of primary astrocytes with GO boosts their homeostatic capacity to control the dynamics of extracellular $K^+$ and glutamate.

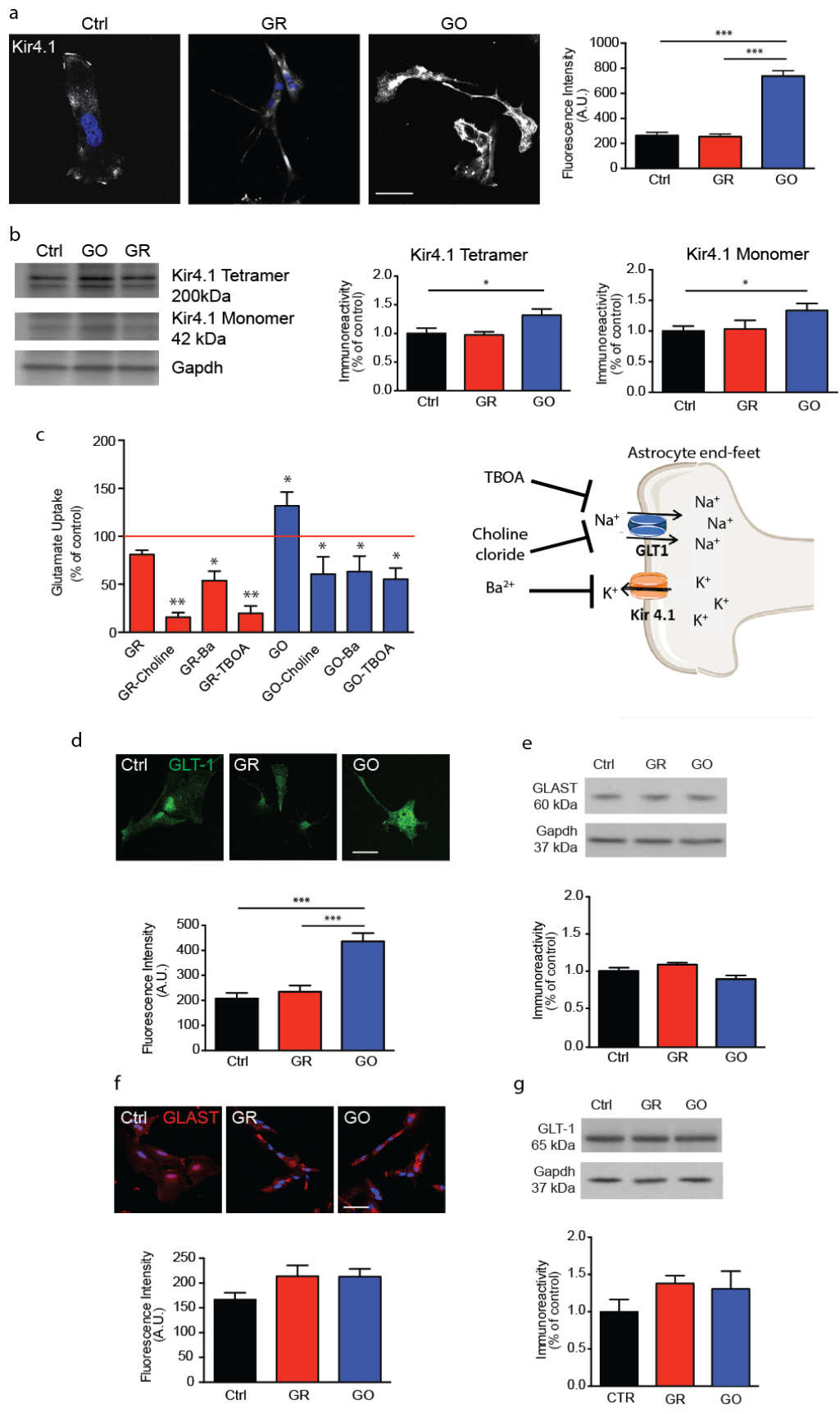

**Figure 3. Kir4.1 expression and glutamate uptake are enhanced in astrocytes exposed to GO. (a)**

Kir4.1 cell surface expression. *Left:* Representative images of primary astrocytes exposed to either GR or GO (10 μg/ml) for 72 h and stained with anti-Kir4.1 antibody (white) and Hoechst33342 (blue). Scale bar, 20 μm. *Right:* Quantitative analysis of Kir4.1 immunoreactivity (n = 40 cells from n = 2 coverslips per experimental condition from 2 independent cell preparations). (**b**) Kir4.1 total expression. *Left:* A representative western blot experiment is shown, in which the tetrameric and monomeric forms of Kir4.1 are electrophoretically resolved. *Right:* Quantitative analysis of the immunoblots shows the overexpression of both the tetrameric and monomeric forms of Kir4.1 (n = 3 from 3 independent cell preparations). (**c**) Glutamate uptake is enhanced in GO-treated astrocytes. *Left:* Quantitative analysis of glutamate uptake by astrocytes was analyzed after 72 h of exposure to either GR or GO (10 μg/ml). Data were normalized to the mean value of untreated astrocytes (n = 3 independent preparations). *Right:* Schematics of the glutamate uptake assay showing the site of action of the inhibitors used. (**d**) GLT-1 cell surface expression. Representative images (*upper panel*) and quantitative analysis (*lower panel*) of GLT-1 immunoreactivity (green) showing an increase surface expression of the transporter upon GO exposure (n = 30 cells, from 2 independent cell preparations). Scale bar, 20 μm. (**e**) GLT-1 total expression. Representative western blot of total GLT-1 (*upper panel*) and quantitative analysis (*lower panel*), showing no changes in the total levels of GLT-1 upon GR or GO exposure. (**f**) GLAST cell surface expression. Representative images (*upper panel*) and quantification (*lower panel*) of GLAST immunoreactivity (red) (n = 30 cells, from 2 independent cell preparations). Scale bar, 20 μm. (**g**) GLAST total expression. Representative western blot of total GLAST and quantitative analysis (lower panels). No changes are observed in both cell surface and total expression of GLAST upon GR or GO exposure (n = 3 from 3 independent cell preparations). In quantitative western blot experiments Gadph was used as loading control. Data are expressed as mean ± sem. *$p<0.05$, **$p<0.01$, ***$p<0.001$, One-way ANOVA/Bonferroni's tests.

### *GO internalization is responsible for shape change, Kir4.1 channel and GLT-1 transporter upregulation*

The effects of GR/GO on the shape and electrical properties of primary astrocytes could be in principle attributable to the membrane shear stress induced by macro-flakes adhering to the external surface of the cell or to the active internalization of nanosheets. In order to discern between these two possibilities and find a mechanistic explanation for the observed effects, we examined the effects of G exposure under conditions in which G internalization was prevented. Primary astrocytes were exposed to either GR or GO for 72 h in the presence or absence of low doses of sodium azide (NaN$_3$; 1 μg/ml), a non-specific blocker of ATP-dependent endocytosis.[39] Astrocytes were then either processed for immunofluorescence analysis (**Figure 4a-e**) or subjected to western blotting for Kir4.1 (**Figure 4f** and **Figure S5**). While the NaN$_3$ treatment *per se* did not alter the cell shape or the viability with respect to vehicle-treated astrocytes, it virtually abolished the marked changes in shape and circularity index induced by either GR or GO (**Figure 4b,d**), likely due to the reduced flake internalization (**Figure 4c,e**). Additionally, NaN$_3$ restored the physiological expression levels of Kir4.1 (**Figure 4f**) and membrane-targeted GLT-1 (**Figure 4g**). As a non-specific blocker of ATP-dependent endocytosis, NaN$_3$ might affect the membrane recycling of receptors and transporters independently of the presence of G flakes. We could however exclude any non-specific effect of NaN$_3$ at the concentration and exposure time used in our experiments, as the expression of Kir4.1 and GLT-1 upon NaN$_3$ treatment was comparable to that of non-treated samples (**Figure 4f,g**).

Furthermore, treatment with NaN$_3$ reverted the GO-induced increase of K$^+$ currents and resting membrane potential, which recovered to the levels of control cells (**Figure 4h-j**). No changes were noticed in Kir4.1 expression upon GR exposure, with or without NaN$_3$ pretreatment (**Figure S5**). Collectively, these data directly link GO internalization to the observed changes in astrocyte shape and plasma membrane K$^+$ conductance, suggesting that the uptake of the material is needed for triggering the observed effects.

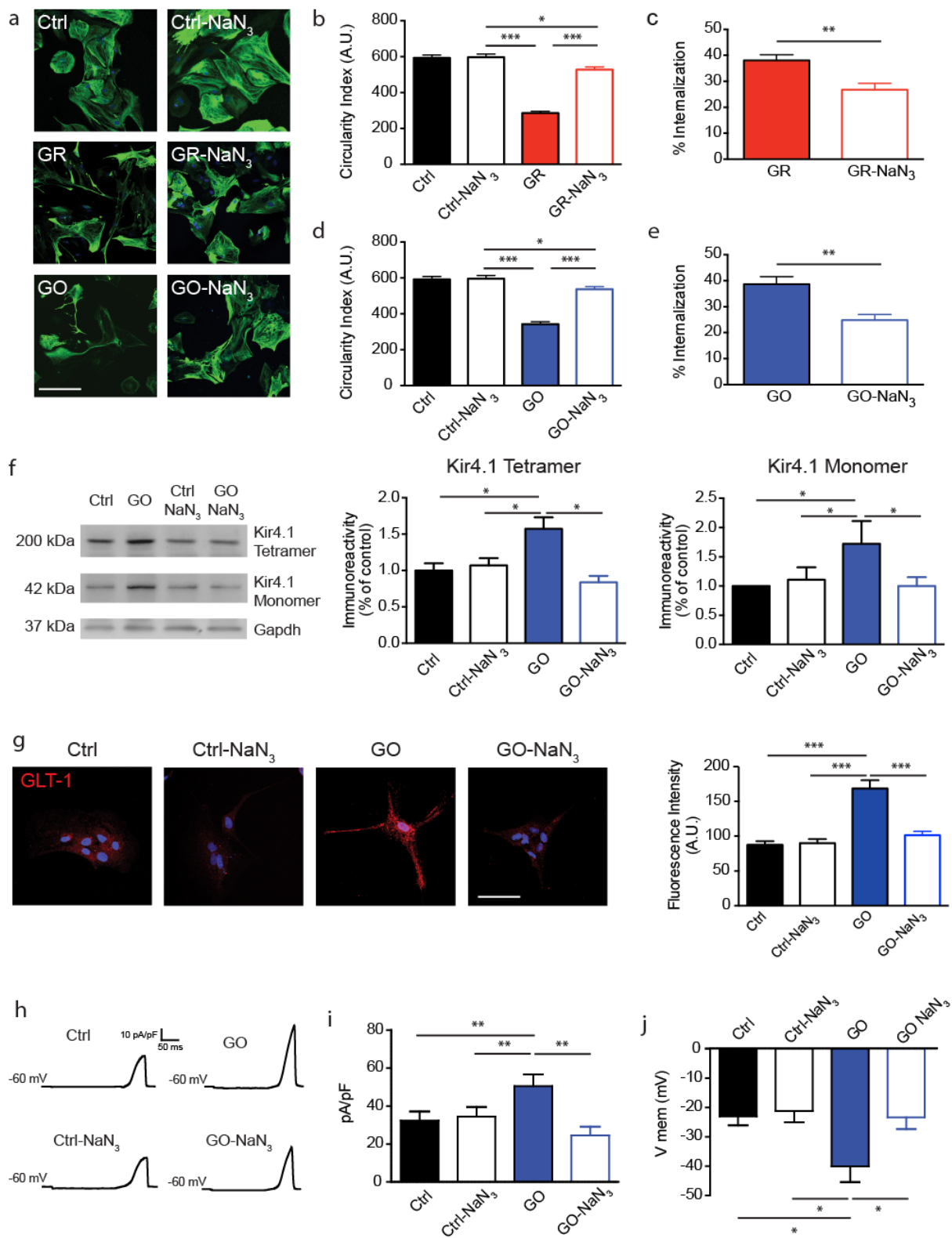

**Figure 4. Blockade of endocytosis by NaN$_3$ rescues astrocyte shape changes and Kir4.1 and GLT-1 increases induced by GO.** Astrocytes were exposed to either GR or GO (10 μg/ml) for 72 h in the presence or absence of 1 μg/ml NaN$_3$. (**a**) Representative images of astrocytes stained for GFAP (green) and Hoechst33342 (blue). Scale bar, 50 μm. (**b,d**) The circularity index was calculated as a measure of the shape change of cells after GR (**b**) or GO (**d**) exposure. (**c,e**) The amount of internalization of GR (**c**) and GO (**e**) flakes was also quantified (n = 100 cells from n = 2 coverslips

per experimental condition, from 2 independent cell preparations). (**f**) Expression of Kir4.1 upon GO exposure in the presence or absence of NaN$_3$. *Left:* A representative immunoblot is shown with the resolved tetrameric and monomeric species of Kir4.1. Gapdh was used as loading control. *Right:* Densitometric analysis of the immunoblots shows that the increased levels of both tetrameric and monomeric species of Kir4.1 are rescued upon NaN$_3$ (n = 3 from 3 independent cell preparations). (**g**) Surface expression of GLT-1. *Left:* Representative images of astrocytes treated with GO (10 μg/ml) for 72 h in the presence or absence of NaN$_3$ and surface stained for GLT-1 (red) and Hoechst33342 (blue). Scale bar, 20 μm. *Right:* GLT-1 membrane expression was quantified (n = 100 cells from n = 2 coverslips per experimental condition, from 2 independent cell preparations). (**h-j**) Whole-cell patch clamp recordings of primary astrocytes exposed to GO in the presence or absence of NaN$_3$. (**h**) Representative current traces elicited with a voltage ramp protocol. (**i,j**) Values of current density at 120 mV (**i**) and resting membrane potential (**j**) under the different experimental conditions (n ≥ 8 cells per experimental condition, from 3 independent cell preparations). Data are expressed as mean ± sem. Panels **c,e**: **p<0.01, two-tailed Student's *t*-test. All other panels: *p<0.05, **p<0.01, ***p<0.001, two-way ANOVA/Bonferroni's tests.

## *GO-exposed astrocytes alter inhibitory synaptic transmission and intrinsic excitability of co-cultured primary cortical neurons*

Potassium buffering and glutamate clearance are fundamental processes in astrocyte-to-neuron communication. Since the above-reported results indicate that G-flakes alter astrocyte homeostatic function, we next investigated whether these changes could affect the *in vitro* development and maturation of primary neurons in astrocyte-neuron co-cultures.

Astrocytes, which had been treated with either GR or GO for 72 h, were used as feeder substrate on top of which, following GR/GO removal, primary neurons in GR/GO-free medium were seeded and grown up to 10 DIV. Under these experimental conditions, GR or GO particles could only be observed in astrocytes and not in neurons (**Figure S6**). Primary cortical neurons were then analyzed at 10 DIV for miniature excitatory (mEPSCs) and inhibitory (mIPSCs) postsynaptic currents and intrinsic excitability. We found that frequency, amplitude, charge and current rise/decay times of mEPSCs did not significantly differ from control co-cultures (**Figure 5a-c** and **Figure S7).** These results were supported by immunostaining co-cultures with antibodies to the vesicular glutamate transporter VGLUT1, which revealed no difference in the density of excitatory synaptic contacts (**Figure 5d** and **Figure S8**). In contrast, a significant increase in mIPSCs frequency was observed in neural networks grown onto GO-treated astrocytes in comparison with GR-treated and control astrocytes, in the absence of changes in the amplitude, charge and current rise/decay times (**Figure 5e-g** and **Figure S7**). To ascertain whether the changes in mIPSCs frequency were due to a concomitant change in the density of inhibitory synapses, we performed immunostaining with antibodies to the vesicular GABA transporter VGAT that revealed a significantly higher density of inhibitory synaptic contacts in neuronal networks co-cultured with GO-pretreated astrocytes with respect to GR-pretreated or control astrocytes (**Figure 5h** and **Figure S8**). Altogether, these observations indicate that exposure to GO enhances the ability of astrocytes to promote the formation and maturation of GABAergic synapses, with no effects on glutamatergic synapses.

Next, we analyzed the intrinsic excitability of co-cultured neurons, a measure of the maturation of voltage-dependent conductance in developing networks. Electrophysiological measurements were performed in current-clamp configuration by injecting 500 ms current pulses with increments of 10 pA, and measuring the resulting action potential firing rate. The analysis of firing rate *versus* injected current revealed that neurons grown on GO-treated astrocytes had a significantly enhanced average and instantaneous firing rates evoked by depolarizing current pulses (**Figure 5i**).

To better analyze the shape of the action potentials, we generated phase-plot graphs by plotting the first derivative of the membrane potential over time (d$V$/d$t$) *versus* the membrane potential. The first component of the phase plot is due to the spike initiation in the axon initial segment (AIS) and to its fast antidromic propagation to the soma that generates a sudden voltage increase from baseline called '*kink*'.[40] The invasion of the soma by the AIS spike generates a delayed activation of somatodendritic $Na^+$ channels that leads to the second phase-plot component, the somatodendritic spike. Interestingly, phase-plots revealed a more prominent '*kink*' in neurons co-cultured with GO-treated astrocytes. To quantify this effect, a linear regression fit of the first 10 experimental points that follow the threshold, set at 5 mV/ms, was performed.[40] This analysis showed that the phase-slope was significantly increased in neurons in contact with GO-treated astrocytes (**Figure 5j**).

These results suggest that GO-treated astrocytes induce an acceleration of the maturation of the intrinsic excitability of cortical neurons by affecting voltage-gated channels not only in the cell soma, but also at the AIS. Although the precise mechanism(s) underlying the observed effects are still not defined, we can envisage at least two possible scenarios, which are not mutually exclusive. On one side, GO could induce long-lasting effects on astrocytes, which would continuously affect neurons up to DIV10 of co-culture. On the other side, GO exposure could affect astrocytes only transiently, in an early time window corresponding to the critical stages of neuronal differentiation and process outgrowth. This early-stage interaction could equally impact on network maturation and synapse formation, leading to the altered electrical properties observed at later stages in the mature network.

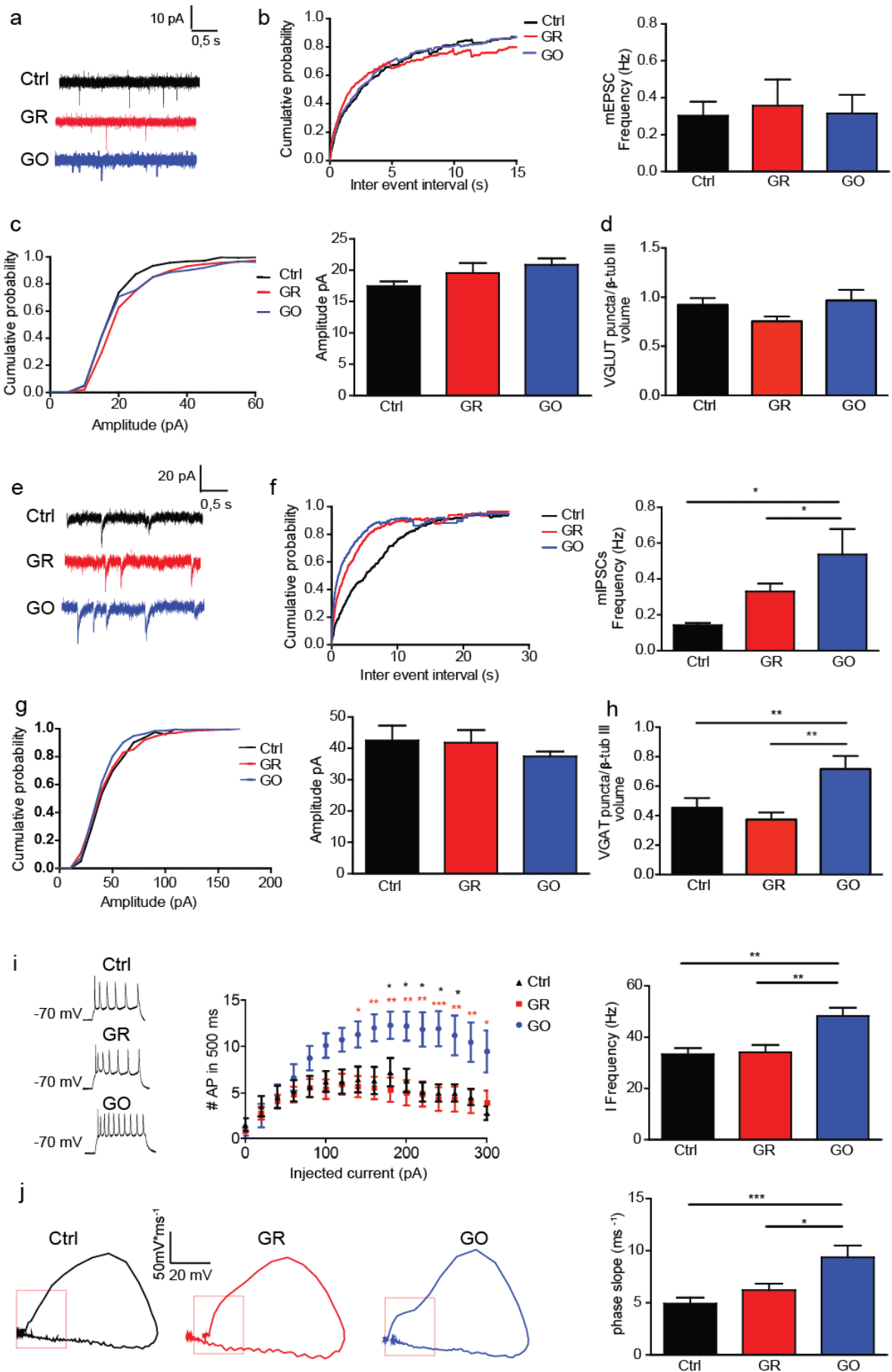

**Figure 5. GO-primed astrocytes affect spontaneous excitatory synaptic currents and intrinsic excitability of co-cultured primary neurons.** (**a-d**) Excitatory transmission. Representative recordings (**a**), cumulative distribution of interevent intervals and frequency (**b**), cumulative distribution of amplitude and mean amplitude (**c**) of mEPSCs (n = 10 cells from Ctrl, n = 10 cells for GR, n = 15 cells for GO) and density of VGLUT1-positive excitatory synapses (**d**; n = 30 cells per experimental condition). (**e-h**) Inhibitory transmission. Representative recordings (**e**), cumulative distribution of interevent intervals and frequency (**f**), cumulative distribution of amplitude and mean amplitude (**g**) of mIPSCs (n = 10 cells per experimental condition) and density of VGAT-positive inhibitory synapses (**h**; n = 30 cells per experimental condition). (**i**) Intrinsic excitability. *Left:* Representative traces of action potentials induced by the injection of 200 pA current steps for 500 ms for the three experimental conditions. *Middle:* Plot of the firing frequency *versus* injected current. *Right:* Instantaneous firing frequency (n = 21 cells for Ctrl, n = 15 cells for GR, n = 25 cells for GO). (**j**) Phase-plot analysis of action potentials. *Left:* Representative plots of the first derivative of the membrane voltage (d$V$/d$t$) *versus* membrane voltage ($V_m$) for each experimental condition. *Right:* Phase-slope measured at voltages more positive of 5 mV/ms (n = 21 cells for Ctrl, n = 15 cells for GR, n = 25 cells for GO). Data are collected from 3 independent cell preparations and expressed as mean ± sem. *$p<0.05$, **$p<0.01$, ***$p<0.001$, one-way ANOVA/Bonferroni's tests.

Astrocytes are fundamental for maintaining brain homeostasis and for the control of key neuronal processes such as synaptogenesis[41, 42] and synaptic transmission.[43, 44] There is also clear evidence that alterations of astrocyte capacity to buffer extracellular $K^+$ and glutamate cause changes in neuronal network excitability, which could result in neurological disorders.[31, 45, 46] In this study we performed a detailed functional analysis of the interaction between GR/GO and primary astroglial cells.

The results show that GR/GO do not affect the viability of cultured astrocytes, which is in contrast with the effects caused by these materials in tumour cell lines.[47] Astrocytes avidly internalize GR/GO flakes to a much higher extent than neurons[8] and the endo-lysosomal pathway results the preferential route of GR/GO intracellular trafficking. In contrast to what was previously observed in neurons,[8] the high amount of internalized material does not induce any autophagy response, likely due to the better capacity of astrocytes to cope with external stressors.

The morphological alterations elicited by GR/GO are triggered by ATP-dependent endocytosis of the flakes and by the subsequent rearrangement of microtubules and microfilaments. The latter finding is consistent with previous results showing that GO nanosheets can insert into the inter-strand gap of actin tetramers, leading to structural alteration of actin filaments.[24] At the functional level, the shape change following GR/GO exposure resembles that of astrocytes subjected to long-term treatment with a membrane-permeable analogue of cyclic AMP (cAMP).[16, 28, 48, 49] These astrocytes have long been considered as a model of reactive astrocytes[49] even though recent work suggests that they may reflect a more mature phenotype.[16]

Notably, GO changes the passive membrane properties of astrocytes by upregulating the $K^+$ conductance mediated by Kir4.1 channels, a molecular signature of mature astroglial cells *in vivo*.[9] Kir4.1 channels preferentially localize at cell processes and perivascular endfeets and are responsible for the negative resting membrane potential of astrocytes.[10] It has been shown that the expression of Kir4.1 and the associated membrane hyperpolarization are positively coupled to increased glutamate uptake.[38] Consistently, GO-treated astrocytes display enhanced $Na^+$-dependent glutamate uptake contributed by both Kir4.1 and GLT-1 upregulation. All these observations support the view that the morphological and functional alterations induced by GO are reminiscent of a more mature phenotype.

Astrocytes shape neuronal activity by modulating neuronal maturation and synaptic transmission.[50] Neurons co-cultured with GO-primed astrocytes display accelerated maturation of intrinsic excitability and increased inhibitory tone. It is tempting to speculate that GO-treated astrocytes promote the functional maturation of neurons by increasing the expression of voltage-gated ion channels and the formation and functional maturation of GABAergic synapses, an effect previously reported to be mediated by astrocyte-secreted neurotrophins.[51]

Other studies have described that astrocytes actively interact with different nanomaterials. Mouse cortical astrocytes grown on single-walled carbon nanotube films displayed increased proliferation and de-differentiated morphology, interpreted as a nonreactive phenotype.[26, 52] By contrast, the same material delivered as colloidal solution promoted the stellation of cultured astrocytes with increased GFAP immunoreactivity[53] and glutamate uptake.[54] Cortical astrocytes plated on synthetic polyamide nanofibers acquired a stellate morphology caused by activation of Rho GTPases[55] and promoted neurite outgrowth when co-cultured with neurons, a result that was associated with the increased secretion of growth factors. The same substrate was also shown to reduce the reactive immunological response of cortical astrocytes differentiated by long-term treatment with cAMP.[56]

Our study demonstrates that long-term exposure of primary astrocytes to GR/GO affects several aspects of their homeostatic capacity, with indirect influence on the functional plasticity of co-cultured neuronal cells. Interestingly, most of the effects are specific for GO, which is probably due to its surface charge and higher reactivity with cells. To date, GO is preferred for biomedical applications with respect to GR, because of its higher solubility, stability in biological fluids and the possibility of surface functionalization. Further studies *in vivo* will address the future potential applications of GO in CNS pathologies.

**Corresponding Author Information**

Corresponding authors: Mattia Bramini, Center for Synaptic Neuroscience and Technology, Istituto Italiano di Tecnologia (IIT), Largo Rosanna Benzi 10, 16132 Genova, Italy; mattia.bramini@iit.it; Stefano Ferroni, Department of Pharmacy and Biotechnology, University of Bologna, Via San Donato 19/2, 40127 Bologna, Italy; stefano.ferroni@unibo.it

**Author Contributions**

M.C. performed cell biology, electrophysiology experiments and analysis, M.B. performed cell biology, cell viability, electron and fluorescence microscopy, immunofluorescence experiments and analysis; A.A. and T.B. participated in designing and discussion of the work; A.R. performed western blotting; E.G. helped with immunofluorescence sample preparation and confocal acquisition; E.V. contributed to the synthesis and characterization of graphene material; M.B., F.C. and F.B. conceived the study; M.B., M.C., S.F., F.C. and F.B. conceived the experimental design and contributed to the analysis of the data; M.B., M.C., S.F., F.C. and F.B. wrote the manuscript. All authors have given approval to the final version of the manuscript.

**Notes**

The authors declare no competing financial interest.

**Funding Sources**

We acknowledge financial support from the European Union's Horizon 2020 Research and Innovation Programme under Grant Agreement No. 696656 - Graphene Flagship - Core1 and Grant Agreement No. 785219 - Graphene Flagship - Core2.

**Acknowledgments**

The Electron Microscopy facility members of the Nanophysics department at the Fondazione Istituto Italiano di Tecnologia (IIT, Genova, Italy) are kindly acknowledged for use and assistance with electron imaging. A. Mehilli is gratefully acknowledged for primary cell culture preparations, as well as D. Moruzzo, F. Canu and I. Dallorto for technical and administrative support. The Antolin Group is also acknowledged for providing the commercial material.

**Supporting Information Available:** Supplementary figures (S1-S8) and detailed experimental procedures (Materials and Methods).

# SUPPLEMENTARY MATERIAL

# Graphene oxide upregulates the homeostatic functions of primary astrocytes and modulates astrocyte-to-neuron communication

*Martina Chiacchiaretta[†,§,°], Mattia Bramini[†,°], Anna Rocchi[†], Andrea Armirotti[#], Emanuele Giordano[†],*
*Ester Vázquez[‡], Tiziano Bandiera[#], Stefano Ferroni[§], Fabrizia Cesca[†,*,ϕ] and Fabio Benfenati[†,*,ϕ]*

[†]Center for Synaptic Neuroscience and Technology and Graphene Labs, Istituto Italiano di Tecnologia, 16163 Genova, Italy; [#]Drug Discovery and Development and Graphene Labs, Istituto Italiano di Tecnologia, 16163 Genova, Italy; [‡]Departamento de Química Orgánica, Universidad de Castilla La-Mancha, 13071 Ciudad Real, Spain; [§]Department of Pharmacy and Biotechnology, University of Bologna, 40126 Bologna; *IRCCS Ospedale Policlinico San Martino, Genova, Italy.

°Equal contribution
ϕSenior authors

*Corresponding authors:*
Mattia Bramini, PhD; e-mail: mattia.bramini@iit.it
Stefano Ferroni, PhD; e-mail: stefano.ferroni@unibo.it

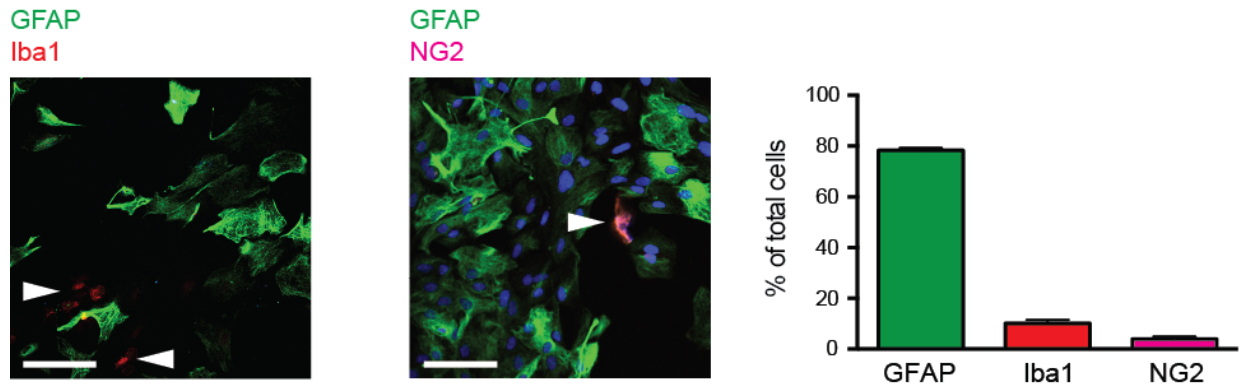

**Figure S1. Characterization of glial cell subpopulations.** *Left:* Glial cell subpopulations were identified by immunofluorescence with anti-GFAP (glial fibrillary acidic protein; astrocytes), anti-Iba1 (ionizing calcium-binding adaptor molecule 1; microglia) and anti-NG2 (neural/glia antigen 2; polydendrocytes) antibodies and counted 72 h after plating. Scale bars, 50 μm. *Right:* Percent composition of the various glial subpopulations (mean ± sem); n = 2000 cells from 3 independent cell preparations.

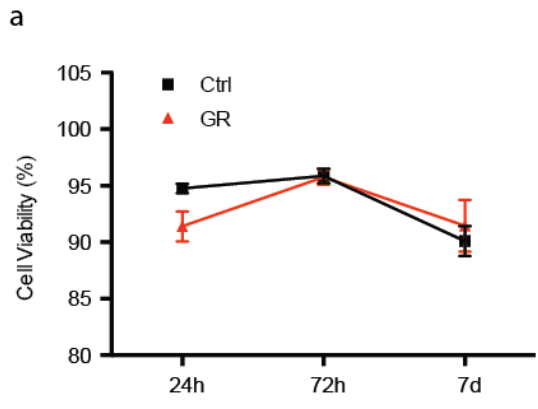
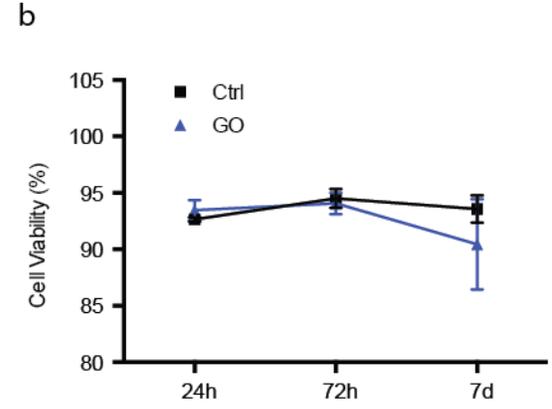
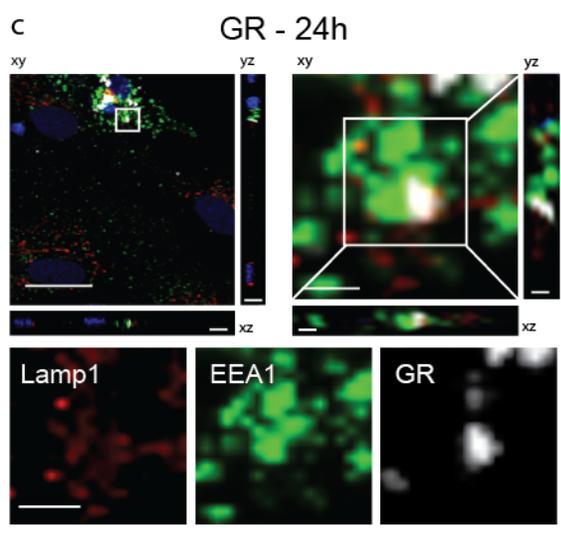
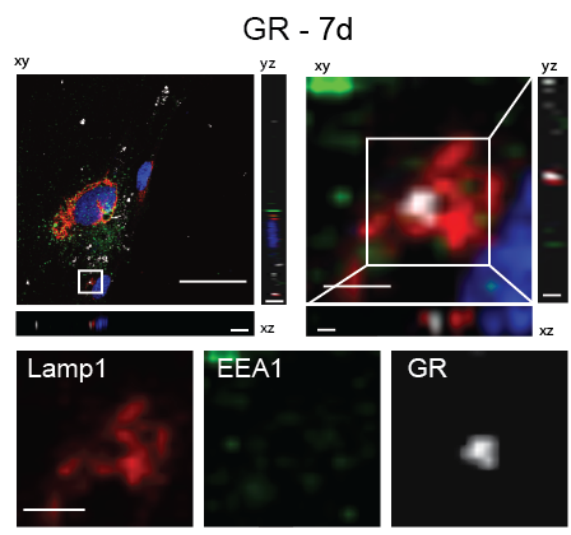
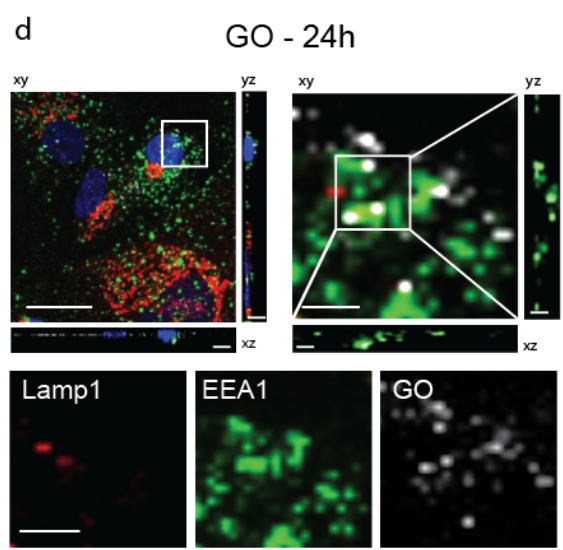
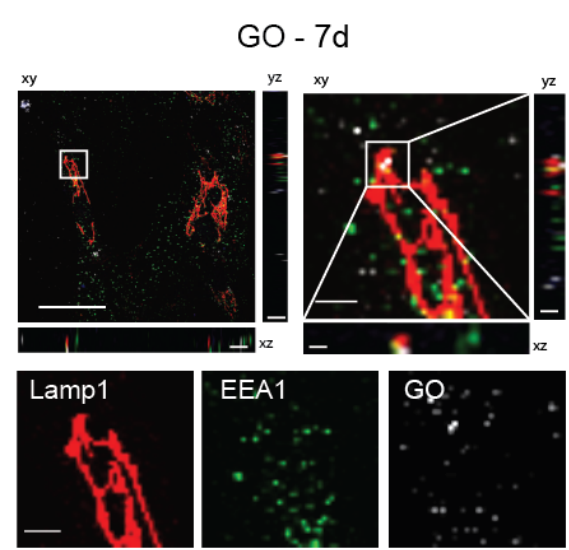

**Figure S2. GR and GO are internalized by astrocytes through the endo-lysosomal pathway.** (**a**,**b**) Astrocytes were exposed to either GR (**a**) or GO (**b**). Cell viability was evaluated by flow cytometry analysis using propidium iodide (PI) staining. The percentages of PI-positive cells with respect to the total number of cells were calculated for each experimental group (45,000 cells from n = 3 per experimental condition, from 3 independent cell preparations). (**c**,**d**) Astrocytes were exposed to either GR (**c**) or GO (**d**) up to 7 days, as indicated. Astrocytes were stained with anti-EEA1 and anti-LAMP1 antibodies for early endosome and lysosome identification, respectively, and Hoechst33342 for cell nuclei. Three-dimensional confocal images were acquired. Scale bars: low magnification xy images (top left of each panel): 20 μm main image and 5 μm xz and yz cross-section; high magnification xy images (top right of each panel): 3 μm main image and 2 μm xz and yz cross-section; single-channel boxes: 3 μm.

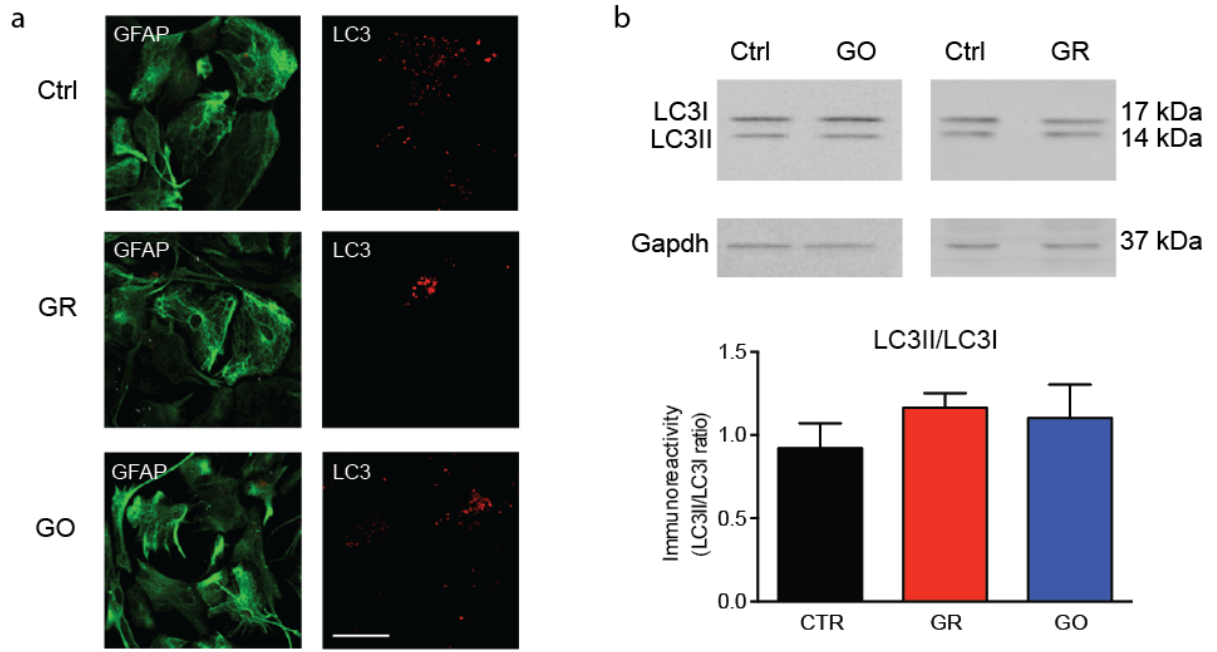

**Figure S3. Internalized GR and GO do not trigger autophagy.** (**a**) Astrocytes were exposed to either GR or GO (10 μg/ml) for 72 h, stained with anti-LC3 and anti-GFAP antibodies, and imaged by confocal microscopy (scale bar, 20 μm). (**b**) Western blotting of LC3 and quantification of the LC3 II/I ratio. Gadph was used as a loading control (means ± sem; n = 3 independent cell preparations).

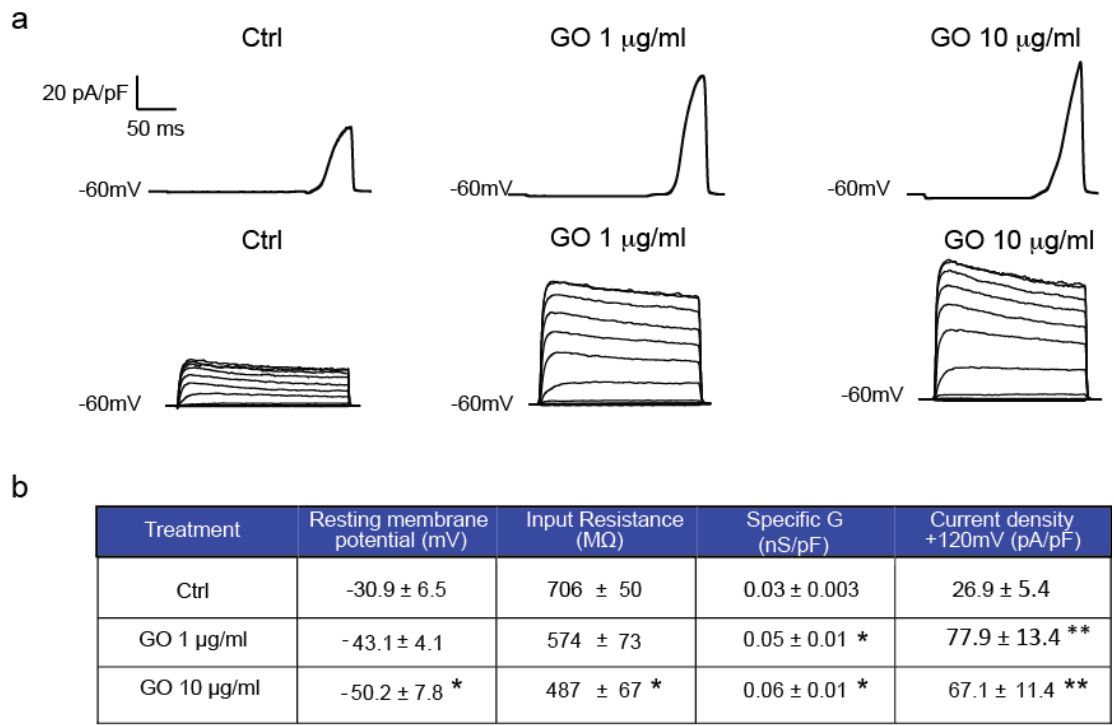

**Figure S4. GO affects the passive membrane properties of astrocytes in a dose-dependent fashion.** (**a**) Representative traces of whole-cell currents evoked with voltage ramps (*upper traces*) or a voltage-step protocol (*lower traces*) in astrocytes treated for 72 h with vehicle, 1 or 10 µg/ml GO. (**b**) Table summarizing the electrophysiological parameters of astrocytes under the various experimental conditions. Means ± sem (n ≥ 10 for each treatment; from 3 independent cell preparations). *p<0.05, **p<0.01, one-way ANOVA/Bonferroni's tests.

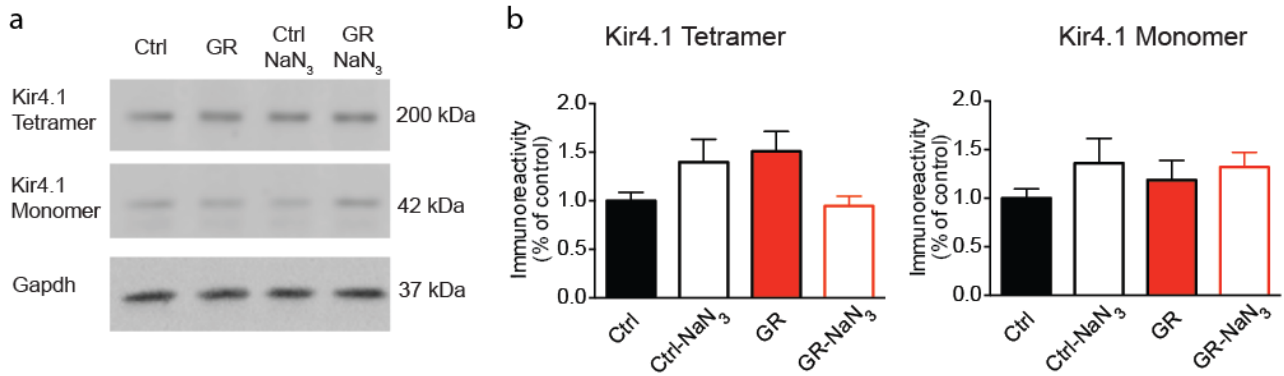

**Figure S5. Kir4.1 expression upon GR exposure and upon blockade of endocytosis.** Astrocytes were exposed to GR (10 μg/ml) for 72 h in the presence or absence of 1 μg/ml of NaN$_3$. (**a**) A representative immunoblot is shown with the resolved tetrameric and monomeric species of Kir4.1. Gapdh was used as loading control. (**b**) Densitometric analysis (means ± sem) shows no significant changes in both tetrameric and monomeric species of Kir4.1 in control and NaN$_3$-treated samples (one-way ANOVA/Bonferroni's tests; n = 2 from 2 independent cell preparations).

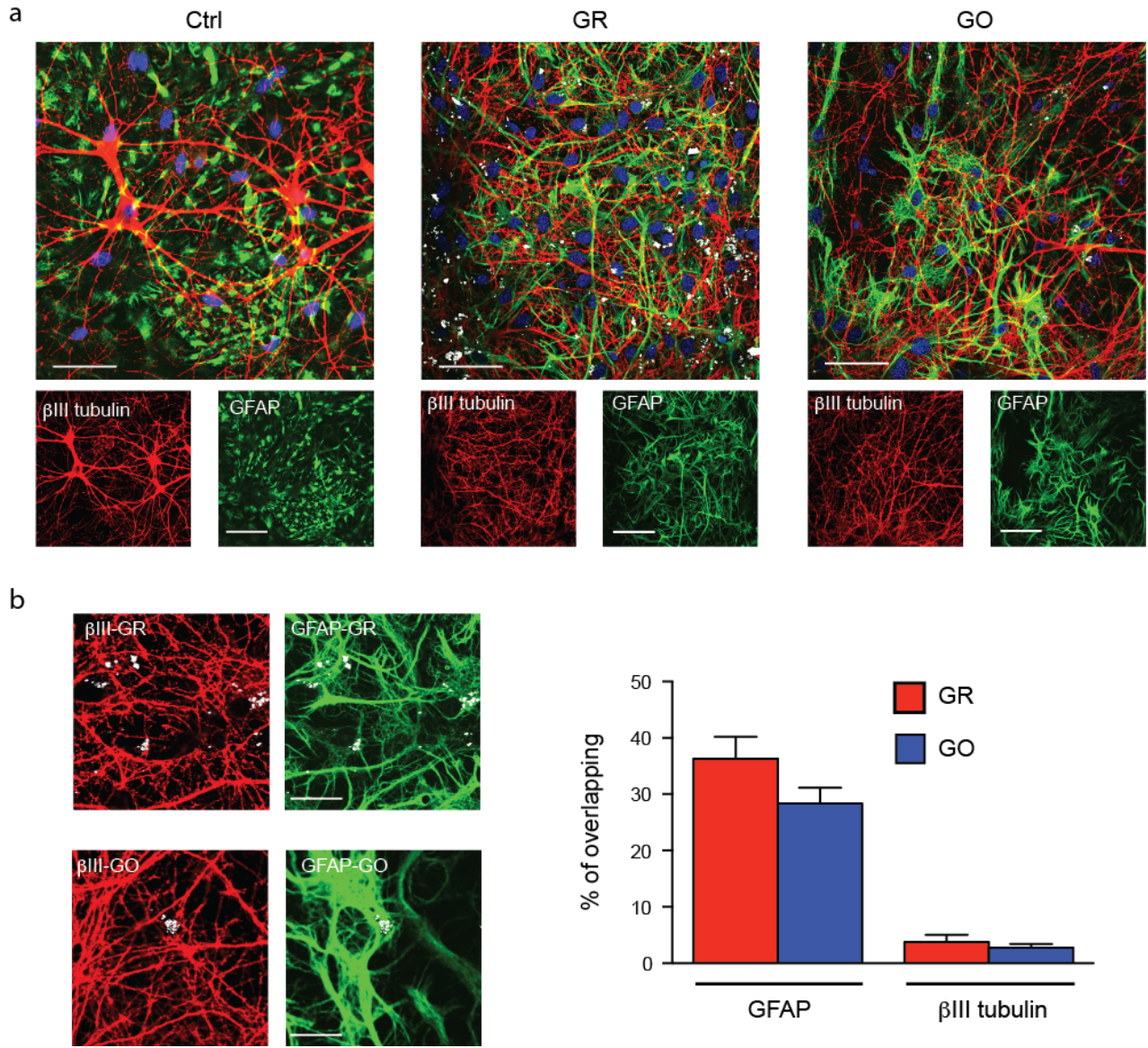

**Figure S6. G flakes are restricted to astrocytes in astrocyte-neuron co-cultures.** Astrocytes that had been pretreated with vehicle (Ctrl), GR or GO (10 µg/ml) for 72 h and washed were used as feeder substrate to plate primary neurons. Neuronal development was then followed in GR/GO-free medium for 10 DIV. (**a**) Representative images of neurons stained for βIII tubulin (red) and astrocytes stained for GFAP (green) at DIV 10. Scale bars, 50 µm. (**b**) *Left:* Representative zoomed images show that GR and GO flakes (white) are predominantly in contact with astrocytes (scale bars: 20 µm for GR and 10 µm for GO). *Right:* quantification of G flakes in contact with astrocytes and neurons, as quantified by reflectance confocal microscopy in double-labeled co-cultures at 10 DIV.

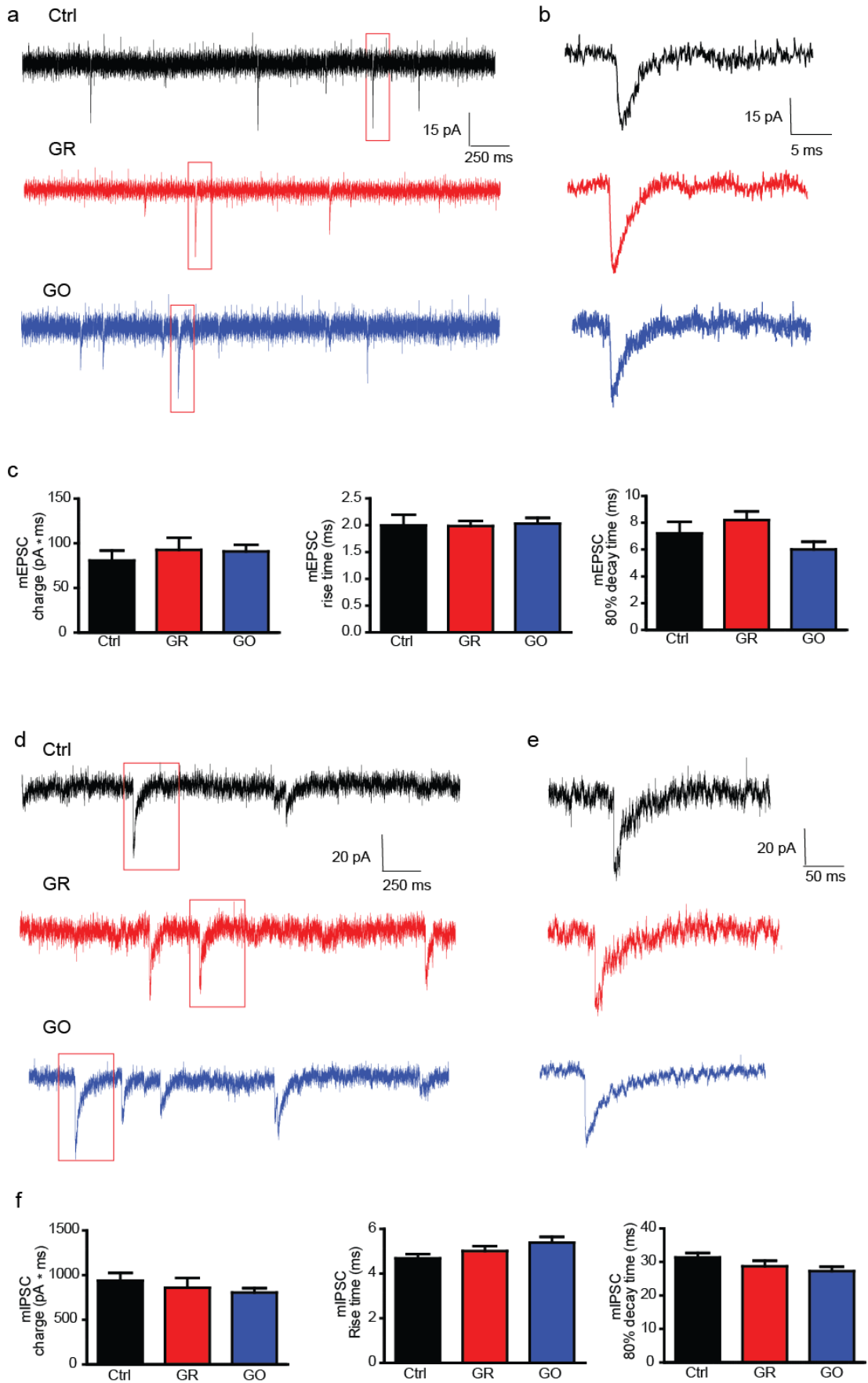

**Figure S7. Exposure to GR or GO does not affect the kinetics of mEPSCs and mIPSCs.** (**a,b**) Representative recordings of mEPSCs (**a**) and examples of individual events over an enlarged timescale (**b**). (**c**) Mean ± sem of mEPSC charge (*left*), rise (*middle*) and decay (*right*) times (Ctrl, n = 10 cells; GR, n = 10 cells; GO, n = 15 cells; from 3 independent cell preparations). (**d,e**) Representative recordings of mIPSCs (**d**) and examples of individual events over an enlarged time scale (**e**). (**f**) Mean ± sem of mIPSC charge (*left*), rise (*middle*) and decay (*right*) times. No significant differences were observed (n = 10 cells per experimental condition from 3 independent cell preparations).

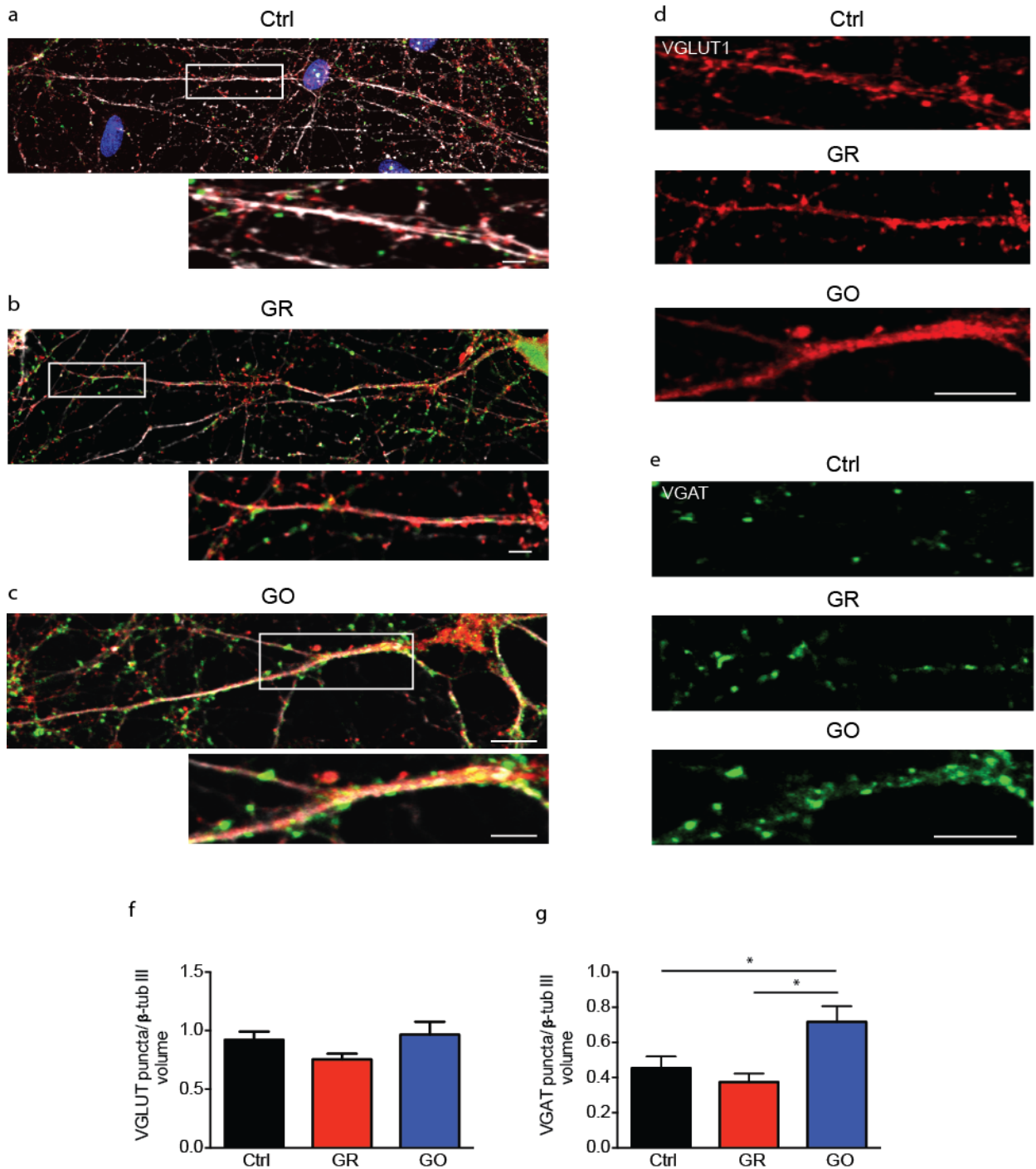

**Figure S8. GO exposure increases the density of inhibitory, but not of excitatory, synapses.** (**a-e**) Representative images of neurons stained for βIII tubulin (gray), VGLUT1 (red) and VGAT (green). Nuclei were stained with Hoechst 33342 (blue). Scale bars: 5 μm in main images (**a-c**) and 2 μm in zoomed images (**d,e**). (**f,g**) Density of VGLUT1-positive excitatory synapses (**f**) and VGAT-positive inhibitory synapses (**g**) in neurons co-cultured with control astrocytes (Ctrl), or with astrocytes that had been exposed to either GR or GO for 72 h before neuron plating. Data are expressed as means ± sem. *p < 0.05, one-way ANOVA/Bonferroni's tests (n = 30 cells per experimental condition, from 3 independent co-culture preparations).

**METHODS**

*Synthesis and characterization of pristine graphene and graphene oxide*

Pristine graphene (GR) flakes were prepared by exfoliation of graphite through interaction with melamine by ball-milling treatment.[12] Graphene oxide (GO) was provided by Grupo Antolin Ingeniería (Burgos, Spain) by oxidation of carbon fibers (GANF Helical-Ribbon Carbon Nanofibres, GANF®) and sodium nitrate in sulfuric acid at 0 °C. Briefly, in GR preparation, melamine was removed by filtration to obtain stable dispersions of few-layer graphene. Elemental analysis found 0.09 ppm melamine in dispersions of 0.09 mg/ml GR; thus the same concentration of melamine was used as control for GR-treated samples. The elemental analysis of GO gave the following results (percent weight): 47.71 ± 0.03 wt % C, 3.04 ± 0.02 wt % H, 0.15 ± 0.01 wt % N, and 0.27 ± 0.03 wt % S. Oxygen content was therefore calculated at ca. 48% by weight. Dynamic light scattering (DLS) and transmission electron microscopy (TEM) analysis displayed a higher lateral size distribution of GR (widely spread from 500 to 2000 nm) compared to GO (100 – 1500 nm, with main peak below 500 nm) flakes. Raman spectroscopy provided a further characterization of the two materials. Graphene exhibits G and 2D modes around 1573 and 2700 cm$^{-1}$, which always satisfy the Raman selection rules, while the D peak around 1345 cm$^{-1}$ required a defect for its activation. For a complete and detailed description of GR and GO material synthesis and physical-chemical characterization, please refer to (Leon *et al.*, 2014)[12] and (Bramini *et al.*, 2016).[7]

*Preparation of primary astrocytes and astrocyte/neuron co-cultures*

All experiments were carried out in accordance with the guidelines established by the European Community Council (Directive 2010/63/EU of 22 September 2010) and were approved by the Italian Ministry of Health (Authorization #306/2016-PR on March 24, 2016). Primary cultures were prepared from wild-type Sprague-Dawley rats (Charles River, Calco, Italy). All efforts were made to minimize suffering and to reduce the number of animals used. Rats were sacrificed by $CO_2$ inhalation, and 18-day embryos (E18) were removed immediately by cesarean section. Briefly, enzymatically dissociated astrocytes were plated on poly-D-lysine-coated (0.01 mg/ml) cell culture flasks and maintained in a humidified incubator with 5% $CO_2$ for 2 weeks. At confluence astrocytes were enzymatically detached using trypsin–EDTA and plated on glass coverslips (Thermo-Fischer Scientific, Waltham, MA) at a density of 20,000-40,000 cells/ml, depending on the experiment. Cultures were incubated at 37 °C, 5% $CO_2$, 90% humidity in medium consisting of DMEM (Gibco/Thermo-Fischer Scientific) supplemented to reach the final concentration of 5% glutamine, 5% penicillin/streptomycin and 10% Fetal Bovine Serum (FBS; Gibco/Thermo-Fischer Scientific). For experiments involving G treatments, cultures were incubated at 1 day *in vitro* (DIV) in a medium containing either 1 or 10 µg/ml of either GR or GO. Controls were subjected to the same medium change with the addition of equivalent volumes of the respective vehicle (0.09 ppm melamine/$H_2O$ for GR flakes, $H_2O$ for GO flakes). Cultures were used at DIV 2, 4 and 8 (i.e., after 1, 3 and 7 days of GR/GO incubation, respectively). For experiments involving astrocyte/neuron co-cultures, astrocytes were seeded on poly-D-lysine-coated (0.01 mg/ml) glass coverslips at a density of 40,000 cells/ml, and incubated for 72 h of incubation with GR, GO or vehicle. After incubation, non-internalized flakes were removed by two washes. Subsequently, enzymatically dissociated cortical neurons obtained from 18-day rat embryos (E18) were plated on the top of astrocytes at a density of 40,000 cells/ml in GO/GR-free Neurobasal (Gibco/Thermo-Fischer Scientific) supplemented to reach a final concentration of 5% glutamine, 5% penicillin/streptomycin, and 10% B27 (Gibco/Thermo-Fischer Scientific). All chemicals were purchased from Life Technologies/Thermo-Fischer Scientific unless stated otherwise. All experiments on co-cultures were performed at 10 DIV.

*Cell viability assay*

Primary rat astrocytes were exposed to GR or GO (10 µg/ml), or to equivalent volumes of the respective vehicle (0.09 ppm melamine/$H_2O$ for GR flakes, $H_2O$ for GO flakes) for 24 h, 72 h and 7 days. The medium was collected in flow-tubes and cells were detached by trypsin EDTA 0.25% (Gibco/Thermo-Fischer Scientific) and re-suspended in 500 ml in PBS in the same flow-tubes of the medium. Cells were stained with propidium iodide (PI, 1 µM) for apoptosis quantification. Cells were incubated at room temperature (RT) for 15 min in the dark. Cells incubated for 8 h with staurosporin (1 µM; Sigma Aldrich) were used as positive control for cell death (not shown). Cell death assays were performed by flow cytometry analysis using a BD Facs Aria II Cell Sorter (Becton Dickinson, NJ, USA).

*Immunofluorescence staining*

Primary astrocytes were fixed in phosphate-buffered saline (PBS)/4% paraformaldehyde (PFA) for 20 min at RT. Cells were permeabilized with 1% Triton X-100 for 5 min, blocked with 2% fetal bovine serum (FBS) in PBS/0.05% Tween80 for 30 min at RT and incubated with primary antibodies in the same buffer for 45 min. The primary antibodies used were: mouse monoclonal anti-early endosomes (EEA1, #610457, BD), rabbit polyclonal anti-lysosomes (LAMP1, #L1418, Sigma-Aldrich), mouse monoclonal anti-glial fibrillary acidic protein (GFAP, #G3893, Sigma-Aldrich), guinea pig polyclonal anti-vesicular glutamate transporter-1 (VGLUT1, #AB5905, Millipore), rabbit polyclonal anti-vesicular GABA transporter (VGAT, #131003, Synaptic System), rabbit polyclonal anti-microtubule-associated protein 1A/1B-light chain 3 (LC3, #2775, Cell Signaling Technology), rabbit polyclonal anti-astrocyte potassium channel KIR4.1 (KIR4.1, #APC-035, Alomone), rabbit polyclonal anti-NG2 (NG2, #AB129051, Abcam), rabbit polyclonal anti-glutamate transporter 1 (EAAT2 – GLT1, #3838, Cell Signaling) and rabbit polyclonal anti-glutamate-aspartate transporter (EAAT1 - GLAST, #AB416, Abcam). For the last two stainings, cells were not permeabilized with Triton X-100 to visualize only the membrane-exposed transporters. After the incubation with primary antibodies and several PBS washes, neurons were incubated for 45 min with the secondary antibodies in blocking buffer solution. Fluorescently conjugated secondary antibodies were from Molecular Probes (Thermo-Fisher Scientific; Alexa Fluor 488 #A11029, Alexa Fluor 568 #A11036, Alexa Fluor 647 #A21450). Samples were mounted in ProLong Gold antifade reagent with DAPI (#P36935, Thermo-Fisher Scientific) on 1.5 mm-thick coverslips. For EEA1, LAMP1, VGLUT1/VGAT-positive terminals and LC3 vesicles, image acquisitions were performed using a confocal microscope (SP8, Leica Microsystems GmbH, Wetzlar, Germany) at 63x (1.4 NA) magnification. Z-stacks were acquired every 300 nm; 10 fields/sample (n = 2 coverslips/sample, from 3 independent culture preparations). Offline analysis was performed using the ImageJ software and the JACoP plugin for co-localization studies. For each set of experiments, all images were acquired using identical exposure settings. For VGLUT and VGAT experiments, values were normalized to the relative cell volume calculated on the basis of β-tubulin III labeling.

*Scanning and Transmission Electron Microscopy*

For SEM analysis, primary astrocytes treated with GR/GO or with the respective vehicle for 72 h were fixed with 1.5% glutaraldehyde in 66 mM sodium cacodylate buffer and post-fixed in 1% $OsO_4$. Sample dehydration was performed by 5 min washes in 30%, 50%, 70%, 80%, 90%, 96% and 100% EtOH solutions. To fully dry the samples, overnight incubation with 99% hexamethyldisilazane (HMDS) reagent (#440191, Sigma-Aldrich) was performed. Before SEM acquisition, coverslips were sputter-coated with a 10 nm layer of 99% gold (Au) nanoparticles in an Ar-filled chamber (Cressington, Sputter Coater 208HR) and imaged using a JEOL JSM-6490LA scanning electron microscope. For TEM analysis, astrocytes treated with GR/GO or with the respective vehicle for 1, 3 and 7 days were fixed with 1.2% glutaraldehyde in 66 mM sodium cacodylate buffer, post-fixed in 1% $OsO_4$, 1.5% $K_4Fe(CN)_6$, 0.1 M sodium cacodylate, *en bloc* stained with 1% uranyl acetate, dehydrated and flat embedded in epoxy resin (Epon 812, TAAB). After baking for 48 h at 60 °C, the glass coverslip was removed from the Epon block by thermal shock using liquid $N_2$. Astrocytes were identified by means of a stereomicroscope, excised from the block and mounted on a cured Epon block for sectioning using an EM UC6 ultramicrotome (Leica Microsystem, Wetzlar, Germany). Ultrathin sections (70 nm thick) were collected on copper mesh grids and observed with a JEM-1011 electron microscope operating at 100 kV and equipped with an ORIUS SC1000 CCD camera (Gatan Inc., Pleasanton, CA). For each experimental condition, at least 6 images were acquired at a magnification up to 10,000x.

*Patch-clamp Electrophysiology*

Astrocytes exposed for 72 h to GR and GO (1 and 10 µg/ml) or to the respective vehicle were used for patch-clamp recordings. The experiments were performed using an EPC-10 amplifier controlled by PatchMaster software (HEKA Elektronik, Lambrecht/Pfalz, Germany) and an inverted DMI6000 microscope (Leica Microsystems GmbH, Wetzlar, Germany). Patch electrodes fabricated from thick borosilicate glasses were pulled to a final resistance of 4–5 MΩ. Recordings with leak current > 100 pA were discarded. All recordings were acquired at 50 kHz. The standard bath saline contained (in mM): 140 NaCl, 4 KCl, 2 $MgCl_2$, 2 $CaCl_2$, 10 HEPES, 5 glucose, pH 7.4, with NaOH and adjusted to 315 mOsm/l with mannitol. The intracellular (pipette) solution was composed of (in mM): 144 KCl, 2 $MgCl_2$, 5 EGTA, 10 HEPES, pH 7.2 with KOH; 300 mOsm/l. Experiments were carried out at RT (20–24°C). All reagents were purchased from Sigma Aldrich. For blocking endocytosis, cells were treated with 1 μg/ml $NaN_3$ (Sigma-Aldrich) for 72 h with GO/GR added 1 h after the start of the $NaN_3$ treatment.

For co-culture experiments, primary rat cortical neurons were seeded on top of astrocytes that had been previously exposed to GR/GO flakes (10 μg/mL) or to the respective vehicle for 72 h, and washed twice in GR/GO-free cell medium. Primary neurons were recorded by patch-clamp experiments after 10 DIV. Recordings of evoked firing activity in current-clamp configuration were performed in Tyrode's extracellular solution in which D-(−)-2-amino-5-phosphonopentanoic acid (AP5, 50 μM), 6-cyano-7 nitroquinoxaline-2,3-dione (CNQX, 10 μM), bicuculline methiodide (BIC, 30 μM), and (2$S$)-3-[[(1$S$)-1-(3,4-Dichlorophenyl)ethyl]amino-2-hydroxypropyl](phenylmethyl) phosphinic acid hydrochloride (CGP, 5 μM) were added to block NMDA, non-NMDA, $GABA_A$, and $GABA_B$ receptors, respectively. The internal solution was composed of (in mM): 126 K gluconate, 4 NaCl, 1 $MgSO_4$, 0.02 $CaCl_2$, 0.1 BAPTA, 15 glucose, 5 Hepes, 3 ATP, and 0.1 GTP, pH 7.3. Current-clamp recordings were performed at a holding potential of −70 mV, and action potential firing was induced by injecting current steps of 10 pA lasting 500 ms. All parameters were analyzed using the Fitmaster (HEKA Elektronik,) and Prism6 (GraphPad Software, Inc.) software. Miniature excitatory postsynaptic currents (mEPSCs) and miniature inhibitory postsynaptic currents (mIPSCs) were recorded in voltage-clamp configuration in the presence of tetrodotoxin (TTX, 300 nM) to block the generation and propagation of spontaneous action potentials. To isolate mEPSC currents, 50 $\mu$M D-AP5 and 30 $\mu$M BIC were added to the extracellular solution to block NMDA and $GABA_A$ receptors, respectively, in the presence of the internal solution (K-gluconate) described above. To isolate mIPSC currents, 10 μM CNQX and 10 $\mu$M CGP were added to the extracellular solution to block non-NMDA and $GABA_B$ receptors, respectively, in presence of an internal solution composed of (in mM): 120 KGluconate, 4 NaCl, 20 KCl, 1 $MgSO_4$, 0.1 EGTA, 15 Glucose, 5 HEPES, 3 ATP, 0.1 GTP (pH 7.2 with KOH). mPSCs were acquired at 10 kHz sample frequency, filtered at half the acquisition rate with an 8-pole low-pass Bessel filter, and analyzed by using the Minianalysis program (Synaptosoft, Leonia, NJ, USA). The amplitude, frequency, rise and decay times of mPSCs were calculated using a peak detector. All reagents were purchased from Sigma Aldrich or Tocris (Tocris, Avonmouth, Bristol, UK).

*Protein Extraction and Western Blotting Analysis*

Total protein lysates were obtained from cells lysed in RIPA buffer (10 mM Tris-HCl, 1 mM EDTA, 0.5 mM EGTA, 1% Triton X-100, 0.1% sodium deoxycholate, 0.1% sodium dodecyl sulfate, 140 mM NaCl) containing protease and phosphatase inhibitor cocktails (Roche, Monza, Italy). The soluble fraction was collected and protein concentration was determined using the BCA Protein Assay Kit (Thermo-Fischer Scientific). For western blotting, protein lysates were denatured at 99 °C in 5X sample buffer (62.5 mM Tris-HCL, pH 6.8, 2% SDS, 25% glycerol, 0.05% bromophenol blue, 5% β-mercaptoethanol, deionized water) and separated on SDS-polyacrylamide gels (SDS-PAGE). The following antibodies were used: mouse monoclonal anti-LC3 (0231, Nanotools, Teningen, Germany), rabbit polyclonal anti-astrocyte potassium channel Kir4.1 (Kir4.1, #APC-035, Alomone), rabbit polyclonal anti-glutamate transporter 1 (EAAT2 – GLT1, #3838, Cell Signaling), rabbit polyclonal anti-glutamate-aspartate transporter (EAAT1 - GLAST, #AB416, Abcam). Signal intensities were quantified using the ChemiDoc MP Imaging System (Biorad, Hercules, CA, USA). For $NaN_3$ experiments, cells were treated with 1 μg/ml $NaN_3$ (Sigma-Aldrich) before incubation with GO/GR.

*Glutamate uptake assay*

For glutamate uptake experiments, astrocytes were plated at concentration of 1 x $10^5$ cell/ml in 6-well plates. Astrocyte cultures were washed three times with pre-warmed (37 °C) HEPES-buffered Hanks' balanced salt solution. 1 μCi/ml of l-[2.3-$^3$H]glutamate (NET490001MC, Perkin Elmer, Milan, Italy) was added to unlabeled l-glutamate (#G8415, Sigma) to reach a final concentration of 50 μM to isolate $Na^+$-dependent transport in standard bath saline contained (in mM): 140 NaCl, 4 KCl, 2 $MgCl_2$, 2 $CaCl_2$, 10 HEPES, 5 glucose. To isolate $Na^+$-independent transport, NaCl was replaced by choline chloride (#C7017, Sigma). To isolate Kir4.1-mediated glutamate uptake, 100 μM $Ba^{2+}$ (#342920, Sigma) was added to the $Na^+$ solution. To block glutamate transporters 100 μM TBOA (#2532, Tocris) was added to the $Na^+$ solution. Cultures were incubated with the isotope on a heating plate at 37 °C for 10 min. After three washes with PBS, cells were harvested into 400 ml of 1 M NaOH solution in $H_2O$. The samples for radioactivity detection were transferred to vials containing 2 ml of aqueous scintillation mixture and counted using a liquid scintillation counter (1450 LSC & Luminescence Counter, MicroBete TriLux, Perkin Elmer).

*Statistical Analysis*

The normal distribution of experimental data was assessed using the Kolmogorov-Smirnov test. Data were analyzed using the Student's *t*-test/Mann-Whitney's *U*-test or, in case of more than 2 experimental groups, by one-way analysis of variance (ANOVA) followed by the Bonferroni's test.